\documentclass[10pt,journal]{IEEEtran}
\usepackage{soul}
\usepackage{times}
\usepackage{verbatim}
\usepackage[usenames,dvipsnames,svgnames,table]{xcolor}
\usepackage[hyphens]{url}
\usepackage[colorlinks=true, linkcolor=blue, urlcolor=blue]{hyperref}
\usepackage{graphicx}
\usepackage{array}
\usepackage{wallpaper}
\usepackage{pifont}
\usepackage{xspace}
\usepackage[longend, ruled, linesnumbered]{algorithm2e}
\usepackage{setspace}
\usepackage{threeparttable}
\usepackage{subcaption}
\usepackage{amsmath}

\usepackage{multirow}
\usepackage{tabularx}
\usepackage{tikz}
\usetikzlibrary{arrows,automata,positioning}

\newcommand{\name}{\textsc{SafeChain}\xspace}
\newcommand{\HIGH}{PRIVATE\xspace}
\newcommand{\LOW}{PUBLIC\xspace}
\renewcommand{\paragraph}[1]{\vspace{1mm}\noindent{\textbf{#1}}}

\graphicspath{{images/}}

\ifCLASSOPTIONcompsoc
  \usepackage[nocompress]{cite}
\else
  \usepackage{cite}
\fi
\ifCLASSINFOpdf
\else
\fi
\begin{document}
%
\title{\name: Securing Trigger-Action Programming from Attack Chains \\ \large{(Extended Technical Report)}}
%
%
%
%

\author{Kai-Hsiang Hsu,
        Yu-Hsi Chiang,
        Hsu-Chun Hsiao,~\IEEEmembership{Member,~IEEE}
        \IEEEcompsocitemizethanks{
          \IEEEcompsocthanksitem K.-H. Hsu, Y.-H. Chiang and H.-C. Hsiao are with the Department
of Computer Science and Information Engineering, National Taiwan University, Taipei, 106, Taiwan.}
\thanks{This is the extended technical report of the 16-page version on IEEE TIFS. Manuscript received April 27, 2018; accepted February 9, 2019.}
}

%
%

\markboth{IEEE TRANSACTIONS ON INFORMATION FORENSICS AND SECURITY,~Vol.~x, No.~x, x~2019}%
{Hsu \MakeLowercase{\textit{et al.}}: SafeChain: Detecting Trigger-Action Attack Chains in Safety-critial IoT}
%



\IEEEtitleabstractindextext{%
\begin{abstract}
  The proliferation of Internet of Things (IoT) is reshaping our lifestyle.
With IoT sensors and devices communicating with each other via
the Internet, people can customize \emph{automation rules} to meet
their needs. Unless carefully defined, however, such rules can easily
become points of security failure as the number of devices and 
complexity of rules increase. Device owners may end up unintentionally
providing access or revealing private information to unauthorized
entities due to complex chain reactions among devices. Prior work on
trigger-action programming either focuses on conflict resolution or
usability issues, or fails to accurately and efficiently detect such
\emph{attack chains}. This paper explores security vulnerabilities
when users have the freedom to customize automation rules using
trigger-action programming. We define two broad classes of
attack---\emph{privilege escalation} and \emph{privacy leakage
}---and present a practical model-checking-based system called \name
that detects hidden attack chains exploiting the combination of
rules. Built upon existing model-checking techniques, \name identifies
attack chains by modeling the IoT ecosystem as a Finite State
Machine. To improve practicability, \name avoids the need to
accurately model an environment by frequently re-checking the
automation rules given the current states, and employs rule-aware
optimizations to further reduce overhead. Our comparative analysis
shows that \name can efficiently and accurately identify attack
chains, and our prototype implementation of \name can verify 100
rules in less than one second with no false positives.

\end{abstract}

}

\maketitle

\IEEEdisplaynontitleabstractindextext

%

\vspace{0.5cm}

\IEEEraisesectionheading{\section{Introduction}\label{sec:introduction}}

\IEEEPARstart{W}{e} now live in an era with smart technologies that
utilize connected devices and sensors to automatically adapt, enhance
performance based on prior experience, and use reasoning to modify the
next behavior~\cite{Mennicken2014}.  According to a recent
survey~\cite{news-gartner}, around 8.4 billion networked devices are
expected to be in use by 2017 and the projected number escalates to
more than 20 billion by 2020.  This speculation indicates that these
Internet of Things (IoT) are already beginning to reshape our daily
lifestyles seamlessly.

Similar to prior advancement in technology, IoT will bring convenience
to our daily lives, at the cost of security and privacy.
As IoT devices are tightly entangled with the physical world, an adversary in cyberspace
can threaten human users' safety and privacy in the physical world via
IoT devices. The lack of appropriate security mechanisms in IoT has
already been highlighted in recent news, ranging from cyber incidents
(e.g., hacking smart fridges to send spam emails~\cite{news-fridge},
compromising smart meters to reduce power bills~\cite{news-meters},
and hijacking toys to leak information~\cite{news-barbie}) to
detrimental cyber-physical threats (e.g., exploiting cardiac devices to
induce inappropriate pacing or shocks~\cite{news-cardiac}, injecting a
worm on IoT devices using ZigBee communication to launch a massive
city-wide light disruptions~\cite{zigbee17}, and compromising IoT
devices to disrupt normal power grid
operations~\cite{blackiot18}).
As more and more vulnerabilities
are discovered, relying on vendors to patch IoT devices in a timely manner is
insufficient. Additional defenses must be in place to limit the impact
on vulnerable devices.


An interesting feature of IoT is supporting customized interaction
among devices using end-user programming, such as trigger-action
programming~\cite{Ur2014}. This often takes the form of ``if trigger,
then action'' and allows users to specify a \emph{trigger} that
represents a condition and the corresponding \emph{action} to be taken
whenever that trigger event occurs. Once defined, such trigger-action
rules can be automatically applied without user involvement. As the
number of connected devices multiplies\footnote{The number of
connected IoT devices per household are anticipated to rise to 50 by
2020~\cite{iot-household}.}, the complexity of interactions among them
will also increase with customized automation. The increasingly
complex interdependencies between devices can easily allow for various
attacks, because an adversary controlling one IoT device can now expand
influence to more devices through such
interdependencies. Unfortunately, attacks leveraging trigger-action
rules are difficult to detect manually, as device owners may
unintentionally provide access or reveal private information to
unauthorized entities due to complex chain reactions~\cite{Wang2018}.

This work presents an automated \emph{prevention} system called \name
which identifies exploitable trigger-action \emph{attack chains}.
\name can thus work in conjunction with methods that support postmortem attack
reconstruction from logs~\cite{Wang2018}, methods that identify errors
in individual rules~\cite{Nandi2016}, and methods that resolve
conflicts between rules~\cite{Liang2015, Liang2016, Ma2017, Ma2016a}.

We first formulate two classes of attack that exploit trigger-action
rules.
The first is \emph{privilege escalation}, in which an
adversary gains control of more devices than it initially has via 
automation rules. For instance, given the rule ``if someone is home,
turn on the light'', an attacker who compromises the occupancy sensor can
also affect the status of the lightbulb.
The other attack class is \emph{privacy leakage}, in which an
adversary learns more information about the devices than it initially
has via automation rules. For example, given the rule ``if someone is
home, turn on the light'', an attacker who observes the state of the
lighting device (e.g., the light is publicly observable or hacked) can
infer the status of the occupancy sensor. In other words, turning the
lighting device on and off with respect to the occupancy of the home
leaks information to the adversary. The attacker can also leverage the
combination of multiple rules to create a chain effect.



To efficiently and accurately identify the two attack classes, we
present \name, a practical system built upon model-checking techniques
and enhanced by domain specific optimizations.
\name models the IoT ecosystem as a Finite State Machine (FSM)\footnote{A
state is a value assignment of all device variables, rules are
modeled as state transitions, and exploitable devices are modeled as
arbitrary change of the variables of these devices at any time.} such
that finding an attack can be reduced to a reachability problem in the
FSM.  Both static and dynamic analysis techniques have been used in
prior work to verify IoT automation.  Static
analysis~\cite{Milijana2017} is often more efficient but comes with
higher false positives as no runtime information is provided.  On the
other hand, prior work~\cite{Yu2015, Liang2016} that similarly
utilizes model-checking tools lacks a clear and detailed
specification, or mostly focuses on resolving conflicts and making sure
individual rules match user intent.  Dynamic analysis,
either FSM or symbolic execution~\cite{Liang2015}, often suffers from
scalability problems and needs a reliable method of modeling the execution
environment.  Therefore, we only consider dynamic analysis to be
practical unless the following challenges have been overcome.


\paragraph{Challenge 1: Environment modeling in FSM.} As model checking
verifies properties against a given ``model,'' an inaccurate model may
 miss detection or create false alarms. Accurately modeling
 environment variables (e.g., trajectory of a user and temperature) is
 nevertheless challenging because it requires extensive knowledge
 about physical environments. Instead of aiming to create an accurate
 environment model (e.g., using differential equations and control
 theory), \name relaxes the requirement by frequently re-calibrating a
 simple environment model based on the current state and the
 extrapolated near-future state. \name then re-checks (e.g., every 1s
 or when the current state changes) the automation rules given the
 updated model.

\paragraph{Challenge 2: The state explosion problem in model checking.}
The number of states in FSM grows exponentially with attributes.
Given hundreds of rules (and device attributes), how can we accurately
and efficiently detect vulnerable rules? In addition, to support
frequent rechecking as stated in the first challenge, the verification
should be able to run as close to real-time as possible. \name employs
two \emph{rule-aware optimization} techniques to reduce redundant
checks and to run significantly faster than using an off-the-shelf
model checker.



Our comparative analysis shows that \name can efficiently and
accurately identify attack chains. Our prototype
implementation of \name can efficiently verify up to 300 automation
rules within one second, outperforming the baseline without any
optimization, which can take more than 15 minutes. The experimental
results also show that \name has no false positives under appropriate
assumptions.


\paragraph{Contributions.}  This paper makes three
contributions:
\begin{itemize}
\item We analyze the attack chains found in a real-world dataset,
  investigate two attack classes
  (i.e., privilege escalation and privacy leakage), and formulate them as checkable
  properties on FSMs.
\item We design and implement \name, a lightweight system to detect
  the two attack classes.
\item We evaluate \name using a large-scale dataset and compare with prior work.  We show that
  \name can verify 300 rules in less than 1s, which is up to 1,000 times faster than the baseline approach, with no false negatives. 
\end{itemize}

\section{Background}
\label{sec:background}
Before explaining how to identify vulnerable trigger-action rules using model
checking, we introduce the related background knowledge and terminologies.



\subsection{Trigger-Action Rules and IFTTT}
Networking capabilities allow IoT devices to communicate and share
information with each other.  For example, an occupancy sensor can
control the lighting or heating system in a smart home when it detects
motion in the space, making daily life more convenient and reducing unnecessary power
consumption. To support such custom automation, users can utilize
trigger-action programming to specify triggering circumstances to
execute actions. The general format of a trigger-action rule is as
follows:

\begin{center}
  IF \emph{trigger}, THEN \emph{action}.
\end{center}
For example, the \emph{trigger} of the previous instance is when
the occupancy sensor detects something and the \emph{action} is to turn on
the light or activate the air conditioner.  Thanks to its
simplicity and straightforwardness, novice users can use such
programming to customize their IoT device behavior~\cite{Ur2014}. 
These rules can also be machine generated~\cite{Mennicken2014,
  Beltagy2016} or learned~\cite{Ur2014, Quirk2015} from user intent.

IFTTT~\cite{IFTTT} is one of the leading services and platforms to
help users define custom automation on their IoT devices.  With more
than one million registered users, IFTTT has
connected more than 400 devices and online services, and in 2015,
more than 19 million rules have been created and 600 million rules
have been executed monthly~\cite{ifttt-qz}.  Other platforms providing
similar services include Samsung
SmartThings~\cite{Samsung-Smartthings}, Zapier~\cite{Zapier} and Microsoft
Flow~\cite{Microsoft-Flow}.


\subsection{Model Checking}
Model checking is a method to formally verify finite-state systems. A
model (i.e., an abstract representation) of a system is automatically
and exhaustively checked to determine if it complies with specified
properties. The desired property of a system is usually expressed in
logic languages, such as Linear Temporal Logic (LTL) and Computational
Tree Logic (CTL).



The characteristics of exhaustive checking from model checking is
especially suitable for security validation, because every hidden
threat can be found with no false alarm. Several off-the-shelf model-checking tools are provided to
help verify systems, such as NuSMV. Once the users model their systems
as finite state machines and express properties in
supported logic languages, the model checker can help determine if there
are any violations.

\section{Case Study}
\label{sec:case}
Before formally defining the problem, we show it is possible to
launch an attack using several harmless automation rules and describe
a home scenario as a working example.

\subsection{Chained Recipes}
\label{ssec:case-studies}
\begin{table}
  \centering
  \caption{Examples of chained recipes}
  \begin{scriptsize}
    \setlength\tabcolsep{3pt}
    \begin{tabularx}{\columnwidth}{|l|X|X|l|}
      \hline
      \textbf{Chain} & \textbf{Recipe 1} & \textbf{Recipe 2} & \textbf{Type} \\
      \hline \hline
      C1 & Convert an e-mail to event in Google Calendar & Send recurring Square Cash payments with Google Calendar \& Gmail & privilege \\
      \hline
      C2 & Disconnect from home Wi-Fi, start recording on Manything & When Manything detects motion, scare the intruder. & privilege \\
      \hline
      C3 & Turn off sprinklers when I arrive home & If irrigation stopped, then blink lights & privacy \\
      \hline
      C4 & When your nest thermostat is set to away then put your water heater in vacation mode & If water heater enters vacation mode, then turn off the lights & privacy \\
      \hline
    \end{tabularx}
  \end{scriptsize}

  \label{tab:chains-for-case-study}
\end{table}
Table~\ref{tab:chains-for-case-study} shows interesting examples of
chained recipes we found that can lead to attacks. The recipes are
all chosen from the public information on the IFTTT
website in April, 2018.\footnote{Because there is no indication
whether a rule is actively used by a user, we assume a user may sign
up for any subset of rules.} For simplicity, we only consider chains
of length two.

The first two examples show privilege escalation attacks. C1 exhibits
how untrusted inputs flow through recipes to a trusted action. The
first recipe of C1 enables almost anyone to create calendar events by
sending mails to the owner, and the second recipe of C1 creates
recurring payments if an event added to Google Calendar matches a
given format. Therefore, an attacker can receive payments from the
victim by simply sending a crafted email. C2 shows another example
  in which the recipes are chained implicitly. The first recipe of
C2 will turn on the security camera (i.e., Manything) when the user
leaves home, while the second recipe of C2 will turn on lights, sounds
or speakers in order to scare an intruder.  This implies that even if
the user is home and the camera is turned off, an attacker can jam the
Wi-Fi to control the house's lighting, sound, or speaker system. We
consider them as chained recipes since the second trigger fires only
after the camera is turned on. Though the attack might not cause any
real damage, it can still be annoying.

The last two examples illustrate privacy leakage attacks, and we
  assume that knowing whether one is home is sensitive, and that lightbulbs
  can be easily observed or compromised~\cite{zigbee17}. With
C3, an attacker can learn that the owner returns home if the
observable light blinks. Similarly, in C4, the light indicates whether
the water heater is in vacation mode, and the water heater's
mode is determined by a nest thermostat. Thus, the thermostat mode
(away or home) is leaked by the light in C4. 

\subsection{A Working Example}
\label{ssec:a-motivating-example}


\begin{figure}
  \centering
  \includegraphics[width=0.6\linewidth]{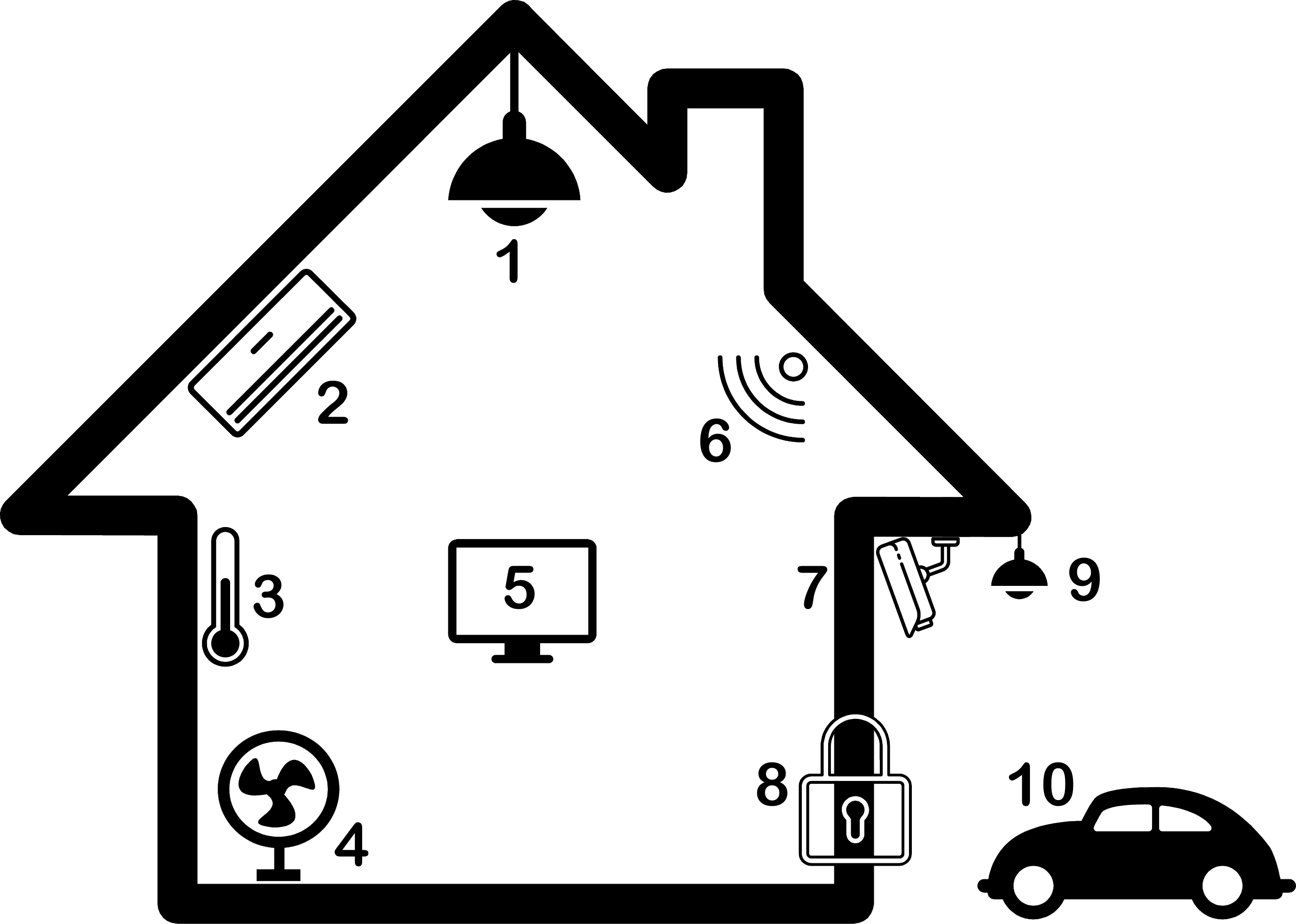}


  \begin{scriptsize}
    \begin{tabular}{|c|c|c|c|}
      \hline
      \multirow{2}{*}{\#} & \multirow{2}{*}{\textbf{Device}} & \multirow{2}{*}{\textbf{Attribute}} & \textbf{Possible} \\
      & & & \textbf{statuses} \\
      \hline \hline
      1 & Light bulb & \texttt{light2} & \textbf{ON}, \textbf{OFF} \\
      \hline
      \multirow{2}{*}{2} & Air & \multirow{2}{*}{\texttt{ac}} & \multirow{2}{*}{\textbf{ON}, \textbf{OFF}} \\
      & conditioner & & \\
      \hline
      3 & Thermometer & \texttt{temperature} & \textbf{0},\textbf{1},...,\textbf{100} \\
      \hline
      4 & Fan & \texttt{fan} & \textbf{ON}, \textbf{OFF} \\
      \hline
      5 & Smart TV & \texttt{tv} & \textbf{ON}, \textbf{OFF} \\
      \hline
      \multirow{2}{*}{6} & Occupancy & \multirow{2}{*}{\texttt{occupancy}} & \textbf{TRUE} \\
      & sensor &  & \textbf{FALSE} \\
      \hline
      \multirow{2}{*}{7} & Surveillance & \multirow{2}{*}{\texttt{camera}} & \multirow{2}{*}{\textbf{ON}, \textbf{OFF}} \\
      & camera & & \\
      \hline
      \multirow{2}{*}{8} & \multirow{2}{*}{Smart lock} & \multirow{2}{*}{\texttt{lock}} & \textbf{LOCKED} \\
      & & & \textbf{UNLOCKED} \\
      \hline
      9 & Light bulb & \texttt{light1} & \textbf{ON}, \textbf{OFF} \\
      \hline
      10 & Smart car & \texttt{location} & \textbf{0},\textbf{1},...,\textbf{127} \\
      \hline
    \end{tabular}
  \end{scriptsize}
  \caption{A Smart Home Scenario}
  \label{fig:home-overview}
\end{figure}

Figure{~\ref{fig:home-overview} illustrates a simplified smart
home scenario consisting of ten devices and 12 rules. Smart home in
reality can be even more complex (and thus harder to analyze), as the
number of connected IoT devices per household are anticipated to rise
to 50 by 2020{~\cite{iot-household}}, and a real dataset provided by a
smart home owner (see Appendix{~\ref{sec:YK}} for details) contains 85
IoT devices connected through nearly 70 automation rules.}  The table
at the bottom of Figure~\ref{fig:home-overview} summarizes the devices
and their possible statuses, and Table~\ref{tab:example-rules} lists
the automation rules used in this example.
Note that rules may be created by multiple users to accommodate individual needs.
A small user study (see Appendix{~\ref{sec:userstudy}} for details) showed that a household of three will have a high
chance to adopt all rules used in this example.

These rules are specified in the format of trigger-action programming
and can work in an automation service like IFTTT. In this example,
the homeowner's intention is to record people entering or leaving the
house. Thus, the surveillance camera will be turned on before the door
is unlocked (R1-R2) and off after the door is locked (R3-R5). The
rules from R6 to R12 are designed for energy efficiency, so the
appliances in the living room will be switched on and off with respect
to different conditions of weather and human presence.

\begin{table}
  \centering
  \caption{Example Rules}
  \resizebox{\linewidth}{!}{
  \begin{scriptsize}
  \begin{tabular}{|c|c|c|}
    \hline
    \textbf{Rule} & \textbf{Trigger} & \textbf{Action} \\
    \hline \hline
    \multirow{3}{*}{R1} & \multirow{2}{*}{User is near home} & Turn on the outside lightbulb \\
    & & Turn on the surveillance camera \\
    & $\texttt{location} = \textbf{0}$ & $\texttt{light1} \leftarrow \textbf{ON}, \texttt{camera} \leftarrow \textbf{ON}$ \\
    \hline
    \multirow{2}{*}{R2} & The surveillance camera is turned on & Unlock the smart lock \\
    & $\texttt{camera} = \textbf{ON}$ & $\texttt{lock} \leftarrow \textbf{UNLOCKED}$ \\
    \hline
    \multirow{2}{*}{R3} & User has been driven out & Lock the smart lock \\
    & $\texttt{location} \neq \textbf{0}$ & $\texttt{lock} \leftarrow \textbf{LOCKED}$ \\
    \hline
    \multirow{2}{*}{R4} & The smart lock is locked & Turn off the outside lightbulb \\
    & $\texttt{lock} = \textbf{LOCKED}$ & $\texttt{light1} \leftarrow \textbf{OFF}$ \\
    \hline
    \multirow{2}{*}{R5} & The outside lightbulb is off & Turn off the surveillance camera \\
    & $\texttt{light1} = \textbf{OFF}$ & $\texttt{camera} \leftarrow \textbf{OFF}$ \\
    \hline
    \multirow{2}{*}{R6} & Occupancy sensor detects someone & Switch on the smart TV \\
    & $\texttt{occupancy} = \textbf{TRUE}$ & $\texttt{tv} \leftarrow \textbf{ON}$ \\
    \hline
    \multirow{2}{*}{R7} & Occupancy sensor detects nobody & Switch off the smart TV \\
    & $\texttt{occupancy} = \textbf{FALSE}$ & $\texttt{tv} \leftarrow \textbf{OFF}$ \\
    \hline
    \multirow{2}{*}{R8} & The smart TV is on & Turn on the inside lightbulb \\
    & $\texttt{tv} = \textbf{ON}$ & $\texttt{light2} \leftarrow \textbf{ON}$ \\
    \hline
    \multirow{2}{*}{R9} & The smart TV is off & Turn off the inside lightbulb \\
    & $\texttt{tv} = \textbf{OFF}$ & $\texttt{light2} \leftarrow \textbf{OFF}$ \\
    \hline
    \multirow{2}{*}{R10} & Temperature is a little high & Turn on the fan \\
    & $\texttt{temperature} \geq \textbf{28}$ & $\texttt{fan} \leftarrow \textbf{ON}$ \\
    \hline
    \multirow{2}{*}{R11} & Temperature is high & Turn on the air conditioner \\
    & $\texttt{temperature} \geq \textbf{32}$ & $\texttt{ac} \leftarrow \textbf{ON}$ \\
    \hline
    \multirow{3}{*}{R12} & \multirow{2}{*}{Temperature is low} & Turn off the fan \\
    & & Turn off the air conditioner \\
    & $\texttt{temperature} \leq \textbf{25}$ & $\texttt{fan} \leftarrow \textbf{OFF}, \texttt{ac} \leftarrow \textbf{OFF}$ \\
    \hline
  \end{tabular}
  \end{scriptsize}}
  \label{tab:example-rules}
\end{table} 

Unfortunately, not every IoT device is equally secure; some might have
vulnerabilities that have not been patched. By compromising a
vulnerable device, an attacker may be able to control or observe other
devices due to inter-device dependencies. In the example in
Table~\ref{tab:example-rules}, suppose the GPS sensor in the smart
car~\cite{news-gps} and the lightbulb~\cite{news-philips} are
compromised, and their states are controlled by an attacker.  The
attacker can infer whether the owner is home (i.e., by knowing the
status of the occupancy sensor), and stealthily break in the house
(i.e., by unlocking the smart lock while the surveillance camera is off)
by leveraging one or a chain of automation rules as we will explain in
\S\ref{ssec:threat-model}.


From the above examples, most of the recipes may look seemingly
harmless if observed individually, but can become harmful when chained
together. As the number of devices and recipes are likely to increase
in the near future, it is harder for humans to debug unsafe chains,
especially when recipes are created at different times or by different
people.

\section{Problem Definition}
\label{sec:problem}



The goal of this work is to create an automated system that can efficiently
detect trigger-action \emph{attack chains} and suggest fixes to
users.
Here, we define an attack chain as a set of rules such that
the action of one rule in the set satisfies the trigger condition of
another rule in the set.

Consider a smart space\footnote{Such a smart space can be a home,
  building, office, factory, etc. For ease of demonstration, we will
  use smart homes (as illustrated in
  \S\ref{ssec:a-motivating-example}) in our examples throughout the
  paper.} consisting of IoT \emph{devices}, user-defined
trigger-action \emph{rules}, and a \emph{service provider}. These devices
can interact with each other explicitly through automation rules.
Such automation is implemented using a trusted
service provider that executes rules whose trigger conditions are
satisfied by the current device state.

In the rest of this section, we define the threat model, system model,
and desired properties in detail.

\subsection{Threat Model}
\label{ssec:threat-model}
We consider a realistic attacker who has compromised a set of
vulnerable devices at the beginning; for example, via malicious apps,
known exploits, or proximity-based attacks~\cite{Wang2018}. The
attacker can read and write attributes of the compromised devices at
any time. Because we are developing a defense system, we consider a strong
  adversary that knows all the rules created by users\footnote{Many
  IFTTT users publish their rules online, and most service providers
  (e.g., IFTTT, Zapier, or Samsung SmartThings) provide a public
  dataset of automation rules so that users can quickly create
  customized automation.}. By successfully mitigating such a
  strong adversary, we can also mitigate weaker adversaries who know partial information,
  as discussed in \S\ref{sec:discussion}.
The attacker's goal is to exploit IoT automation and perform
unauthorized actions \emph{(privilege escalation)} or unauthorized
reads \emph{(privacy leakage)}.

\paragraph{Privilege escalation.}
In a privilege escalation attack, the attacker attempts to make the
IoT system transition into an insecure state (e.g., the door is unlocked
when cameras are off), which can never be reached if the devices are operating
as expected. To do this, the attacker actively manipulates the
attributes of the compromised devices, thereby triggering changes of
other devices via automation rules. Sometimes the attacker may also
need to manipulate the device attributes in a specific sequence and at a specific
time.



In the example in \S\ref{ssec:a-motivating-example}, the attacker can
manipulate the state of the smart lock
and surveillance camera, thus break in stealthily without being
recorded, even with no direct control of the two
devices. To achieve this, the adversary forces the GPS sensor to
incorrectly report (e.g., by generating a stronger, fake GPS signal
or hacking the backend service~\cite{remote-smart-car-hacking})
that the car is home. The service provider is then misled to apply
rules R1 and R2 to turn on the outside lightbulb and surveillance
camera and unlock the smart lock, respectively. After that, the
adversary forges the status of the lightbulb to trigger rule R5,
which turns off the surveillance camera.

\paragraph{Privacy leakage.}
In a privacy leakage attack, the attacker attempts to deduce private
information from publicly observable data and the attributes of
compromised devices. In addition to passive
observation, the attacker can also actively manipulate compromised
devices and observe the resulting changes.



In the example in \S\ref{ssec:a-motivating-example}, the attacker can
infer whether the owner is home by monitoring
the state of the vulnerable lightbulb inside. This is because when the occupancy
sensor detects a human presence, rules R6 and R8 will be triggered, thus
turning on the smart TV and the inside lightbulb.  On the other hand,
if the occupancy sensor detects no one, rules R7 and R9 will be
applied, and both the smart TV and inside lightbulb will both be
off. Thus, the adversary can infer the state of the occupancy sensor
through monitoring the state of the inside lightbulb.

\subsection{System Model}
\label{ssec:system-model}


\paragraph{Devices.}
A device's state can be represented using a set of attributes,
which can be accessed via APIs. For example, a thermometer can have a
\texttt{temperature} attribute, which is set
to the value perceived by its temperature sensor;
a lightbulb or surveillance
camera can have a \texttt{switch} attribute that represents whether
its functionality is enabled or disabled.

Note that these attributes may be affected by the
time-varying \emph{environment} via the devices' sensors. They may
also affect the environment via the devices' actuators.


\paragraph{Trigger-action rules.}
Users can enable automation between devices by adding customized
trigger-action rules.
Users need to specify a trigger and the corresponding
action when creating a customized rule.





\paragraph{A service provider.}
We consider IoT automation implemented using a trusted service
provider\footnote{Securing the service provider is an orthogonal
  problem to our work, but we emphasize that the service provider has
  incentives and resources to employ better security measures than
  individual IoT devices. Many IoT vendors focus on providing novel
  functionalities rather than security, and IoT devices rarely adopt
  strong security measures due to limited resources.  On the other
  hand, once the service provider is compromised, the attacker can
  directly control all devices to launch powerful attacks, which will
  ruin the service provider's reputation.}  (e.g., IFTTT, Zapier, or
Samsung SmartThings).  The service provider offers an interface for
users to add or remove customized rules and can interact with devices
through their APIs. We assume that the service provider
polls devices periodically; that is, queries every device's status to check for
satisfying rules and to apply corresponding actions. We also
assume that the service provider can resolve conflicting rules and
avoid ambiguities, e.g., by enforcing an order of precedence on
rules~\cite{maternaghan2013policy, resendes2014conflict}.





\subsection{Desired Properties}
\label{ssec:desired-properties}
\paragraph{Low false rates.}
The system should accurately identify privilege escalation
and privacy leakage attacks. A false positive occurs when the system
falsely reports an attack, which may annoy users or affect normal
functionality of IoT devices.  A false negative occurs when the
system fails to identify an attack, providing users with a false sense of security.


\paragraph{Timely detection.}
The system should be able to scale to hundreds or even thousands of
devices and rules, and detect potential attacks in a timely
manner. Timely detection ensures that users have sufficient time to fix
problematic rules before they are exploited. Designing a
scalable algorithm is challenging because the number of possible state
combinations 
grows exponentially with devices and rules.


\paragraph{Low interference with intended functionalities.}
One  way to prevent attackers from exploiting automation rules
is to ask users for permission before executing every rule. However,
this simple fix contradicts the purpose of IoT automation, which is to
make users' lives easier. Hence, the mitigation method should avoid
interfering with users' intended functionalities and should not place
a burden on users.



\section{\name}
\label{sections:veriiot}

The core idea of \name is to model the IoT ecosystem as a Finite State
Machine (FSM), such that finding an attack can be reduced to a property-checking problem on the FSM, which can be solved using existing model
checkers. To improve its practicability, \name deploys novel
techniques to overcome the research challenges described in
\S\ref{sec:introduction}.

After highlighting the core insights and the system overview
(\S\ref{ssec:overview}), we
explain how \name models the problem of identifying automation rule
exploitation as a reachability problem on a finite state machine
(\S\ref{ssec:modeling}), such that we can adopt model-checking
techniques to ensure accurate detection of privilege escalation and
privacy leakage (\S\ref{ssec:verification}). As model checking tends
to be a slow process, we propose several rule-aware optimization
techniques to achieve timely detection
(\S\ref{ssec:optimization}). Finally, we explore how to mitigate the
identified attacks with low interference (\S\ref{ssec:mitigation}).

\begin{figure*}
  \centering
  \includegraphics[width=0.8\textwidth]{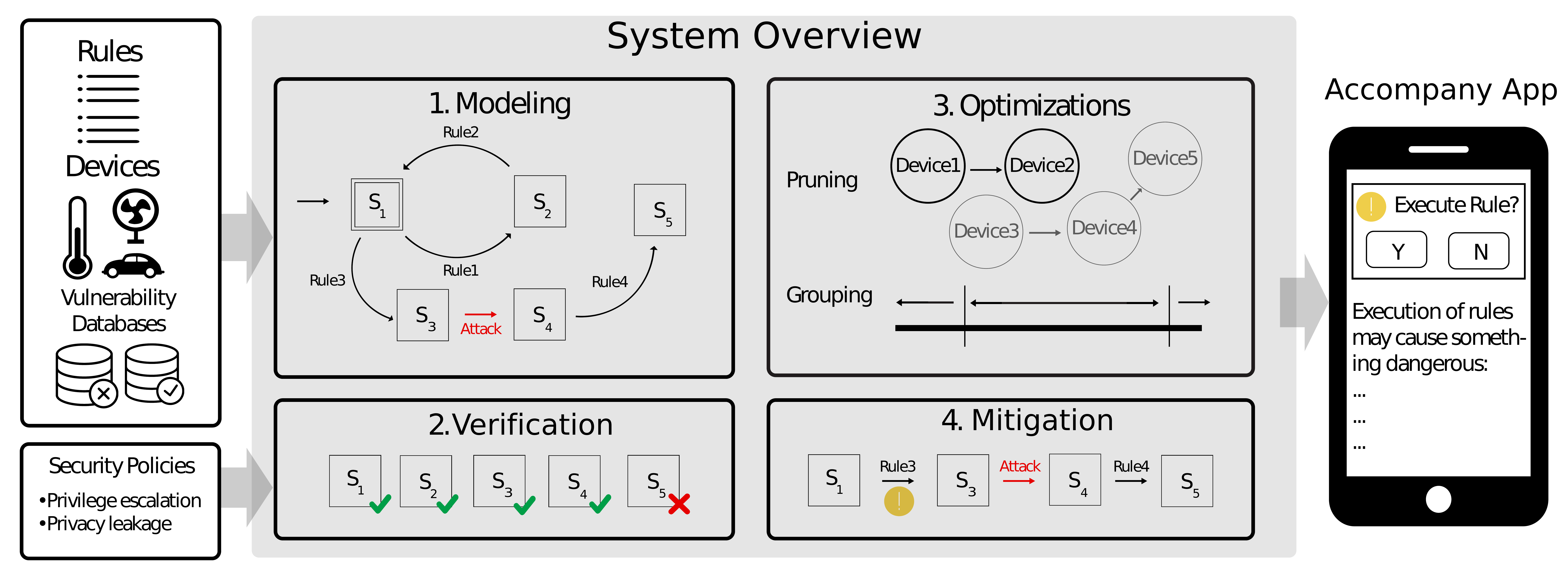}
  \caption{System overview. ``S'' stands for states.}
  \label{fig:overview}
\end{figure*}

\subsection{High-Level System Overview}
\label{ssec:overview}






As shown in Figure~\ref{fig:overview}, \name takes in
 rules, devices, vulnerability databases, and security
policies, and reports identified attacks to users.  Users can
interact with \name using an application interface to check detailed
attack traces and apply fixes. 

We envision that \name can work as an extension to an existing IoT
automation service provider or as a standalone system, and the inputs
can be supplied by the service provider, IoT vendors, crowdsourcing
sites, users, etc. Experts can help provide default security policies,
and users can revise them according to their needs.

Because the adversary's goal is to control an uncompromised device or
leak data from an uncompromised device via automation rules, it is
reasonable to consider cases where some devices are vulnerable and
some are not. (The attacker succeeds immediately if all devices are
compromised.) Several approaches exist to determine the vulnerable
devices. For example, vulnerability databases provide lists of
vulnerable devices\footnote{Several public datasets have been
established to consider the consumer device vulnerabilities. For
instance, National Vulnerability Database~\cite{nvd} and
AndroidVulnerabilities~\cite{avo} have accumulated a variety of
vulnerabilities with respect to different devices and smart phones.}.
Alternatively, users can manually select devices that require high
level of protection, in which case our system assumes that all the
other devices are vulnerable and evaluates whether the
manually-selected devices are attackable.


\name consists of four major components:
\begin{enumerate}
\item \textbf{Modeling.}
  To model the interaction between devices and rules, we build a FSM
  in which the device statuses and automation rules correspond to the
  states and transitions, respectively. In addition, to model a
  volatile environment (Challenge 1), we use a short-lived window to
  predict the changes of each sensor variable and renew the prediction
  after the previous one expires. 

\item \textbf{Verification.}  To verify the system, it is crucial to
  define the attacker model in the FSM. The verification component
  translates the given security policies into the FSM properties, which can be
  checked using a model checker. Once the short-lived windows
  are due, our system re-verifies the model again with the new windows.
\item \textbf{Optimization.}  To scale to a large number of devices
  and rules (Challenge 2), we propose two rule-aware optimization
  techniques to shrink the size of the FSM by pruning redundant states
  and grouping equivalent states. Since the optimization is done before
   transforming into FSM, with high-level semantics preserved,
  our approach incurs less overhead compared to general
  optimization techniques implemented in common model checkers.
\item \textbf{Mitigation.}
The mitigation component greedily selects a small set of rules whose
removal can disable the identified attacks. To avoid violating
intended functionalities, this component also shows the identified
attack traces and suggests fixes to the users.
\end{enumerate}

As shown in \S\ref{sec:experiment}, \name can verify hundreds
of rules in just a few seconds, and thus is capable of frequent
rechecks based on the latest rule set and sensor-attribute
values. Note that frequent rechecks are needed to accommodate the
short-lived window for the modeling environment.

\subsection{Modeling}
\label{ssec:modeling}
We will first describe how we encode each concept separately, and
then explain in detail how \name models the whole smart space to
simulate the possible interactions.
The same modeling is used for detecting both privilege escalation and privacy leak.

\paragraph{Devices.}
Each device is symbolized by using a set of attributes to
represent its equipped sensors, actuators or internal states. 
%
We use $A$ to denote this attribute set of all devices,
which are classified into two disjoint groups: read-only
attributes $A_R$ and read-write attributes $A_{RW}$. The former
corresponds to sensors, which provide APIs only for obtaining values,
and the latter corresponds to actuators, which provide APIs for obtaining
and setting values. Possible values of each attribute $a_i \in A$
are specified in its API specifications and we use $possible(a_i)$ to
denote this set.

\paragraph{Automation rules.}
A rule, represented by $r_i \equiv (bool_i, assign_i)$, can be
specified by users or generated by machines. $bool_i$ is a
Boolean-valued function defined over the device attribute space, and
$r_i$ is triggered when $bool_i$ is evaluated to be \textbf{TRUE}.
$assign_i$ defines the action of the rule. This function maps device attributes to values.




For simplicity, we assume that all the satisfied rules (i.e.,
$bool_i(s) = \textbf{TRUE}$) are executed concurrently. 
We also assume that if there is a conflict between rules, some
resolution techniques are applied, such as user preferences, to ensure
that the service provider knows which rules to execute.  Two rules
$r_i$ and $r_j$ are said to be conflicting if they assign different
values to the same variable (i.e., $assign_i(a_k) \neq assign_j(a_k)$).

\paragraph{Environment.}
The values of sensor attributes will change as the environment
changes and corresponding sensors perceive the differences.
At one extreme, we
could try to model the environment accurately using knowledge and
formula in physics. However, since the smart space can be a home,
office or factory, they may differ greatly in their environments. It
is hard to use one or few formulas to cover all conditions. At the
other extreme, we could set every sensor attribute to be
constraint-free and check every combination of their values at any
time, but this may cause excessive false alarms
because some combinations may never happen in the real world.

To strike a middle ground between the two extremes, we propose a
practical approach to handle environmental changes in \name by
focusing on possible changes in the near future. Specifically, for
each sensor attribute $a_i \in A_R$ of a secure device, we try to
predict a window $window(a_i) \subseteq possible(a_i)$ in which this
attribute will reside during a period of time, and after the period
has elapsed, we repeat the prediction process and verify the rules
again. For example, the temperature is expected to be between 23 and
33 Celsius degrees tomorrow, based on the weather forecast. To ensure that no
attacks will occur tomorrow, there is no need to check values outside
this window (unless the thermometer is assumed to be hacked). A similar
assumption can be made for the GPS sensor in a smart car, because the
car movement obeys the laws of physics and cannot move faster than a
certain speed. For sensors without known constraints on their
attribute values (e.g., an \texttt{occupancy} sensor),
we enumerate all possible conditions in the future, i.e. $window(a_i) =
possible(a_i)$.


In addition, to alleviate the impact of inaccurate prediction, \name
will monitor the sensing data and immediately recheck and re-predict
if the prediction is violated. In our implementation, the prediction
window is derived from a fixed width for each sensor attribute and can be adjusted with respect to users' tolerance to false alarms. It
can also be improved by using techniques such as machine learning.


\paragraph{Adversary.}
Given the vulnerability databases, \name uses $A_{VUL} \subseteq A$ to denote
the attribute set of vulnerable devices.
Any attribute in $A_{VUL}$ can be monitored or modified by an attacker
at any time.



An attacker can be either active or passive. A passive attacker
gathers information about compromised devices over time and tries to
infer information about secure devices, while an active attacker
reports bogus information to trigger automation rules.
We use a special attribute \texttt{attack}, which can be either
\textbf{ACTIVE} or \textbf{PASSIVE} at a time, to represent the chosen
strategy.

The whole smart space is then modeled as a finite state machine, which
is a tuple $FSM \equiv (S, \longrightarrow, I)$ where $S$ is a set of
states, $\longrightarrow \subseteq S \times S$ is a transition
relation, and $I \subseteq S$ is a set of initial states.

A state $s \in S$ is an $N$-tuple $(a_1, a_2, \ldots a_N)$, where
$a_i \in A$ are the attributes of all installed devices. We use the
notation $s(a_i)$ to represent the value of $a_i$ in $s$.
Set $S$ consists of all the possible states in the smart space while
set $I$ contains only one state representing the current status.
The next possible state $s'$ from state $s$ can be affected by automation
rules, environment, and the adversary simultaneously. Formally for
state $s$, the transition relation $(s, s') \in \longrightarrow$
exists if for any $a_i \in A$,
\[
s'(a_i)
\left \{
  \begin{scriptsize}
  \begin{tabular}{ll}
  $\in possible(a_i)$ & if $a_i \in A_{VUL}$ and $s'(\texttt{attack}) = \textbf{ACTIVE}$ \\
  $\in window(a_i)$ & if $a_i \in A_R$ \\
    $= assign_k(a_i)$ & if $bool_k(s) = \textbf{TRUE}$ for some rule $r_k$ \\
  $= s(a_i) $ & otherwise.
  \end{tabular}
  \end{scriptsize}
\right \}.
\]
The first condition corresponds to the case when the attacker actively
controls the devices so that the vulnerable attributes can be set to any
possible value. The second condition limits the environment attributes
to remain in our predicted values, and the third defines the effects
of automation rules. In addition to what is discussed, the status of devices will remain
unchanged.

\subsection{Verification}
\label{ssec:verification}


In this subsection, we explain the formats of security policies
  for privilege escalation and privacy leakage, and how such policies
  are translated into FSM's properties that can be verified using
  model checking tools. We envision that (1) our system has a set of
  pre-installed general security policies written by experts, and (2)
  users can modify or create their own policies, and share their
  policies to a public dataset. Even if sophisticated
  users can write their own security policies, it may still be challenging for
  them to manually check all possible interactions between devices.



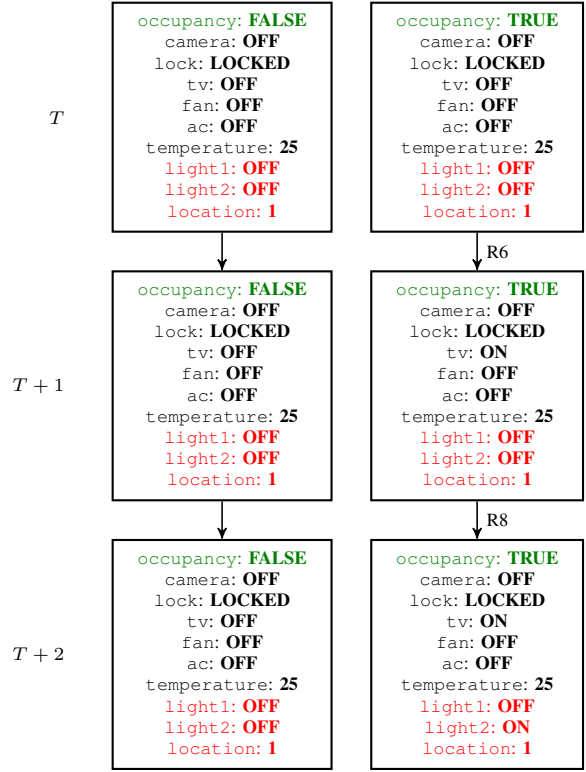
\begin{figure}[t]
  \centering
  \begin{tikzpicture} [->,>=stealth',semithick]
  \tikzstyle{state} = [draw, thick, fill=white, rectangle];
  \node[state] (x0) {\begin{scriptsize}\begin{tabular}{c}
    \textcolor{Green}{\texttt{occupancy}: \textbf{FALSE}} \\
    \texttt{camera}: \textbf{OFF} \\
    \texttt{lock}: \textbf{LOCKED} \\
    \texttt{tv}: \textbf{OFF} \\
    \texttt{fan}: \textbf{OFF} \\
    \texttt{ac}: \textbf{OFF} \\
    \texttt{temperature}: \textbf{25} \\
    \textcolor{Red}{\texttt{light1}: \textbf{OFF}} \\
    \textcolor{Red}{\texttt{light2}: \textbf{OFF}} \\
    \textcolor{Red}{\texttt{location}: \textbf{1}}
  \end{tabular}\end{scriptsize}};
  \node[state] (x1) [below = 0.5cm of x0] {\begin{scriptsize}\begin{tabular}{c}
    \textcolor{Green}{\texttt{occupancy}: \textbf{FALSE}} \\
    \texttt{camera}: \textbf{OFF} \\
    \texttt{lock}: \textbf{LOCKED} \\
    \texttt{tv}: \textbf{OFF} \\
    \texttt{fan}: \textbf{OFF} \\
    \texttt{ac}: \textbf{OFF} \\
    \texttt{temperature}: \textbf{25} \\
    \textcolor{Red}{\texttt{light1}: \textbf{OFF}} \\
    \textcolor{Red}{\texttt{light2}: \textbf{OFF}} \\
    \textcolor{Red}{\texttt{location}: \textbf{1}}
  \end{tabular}\end{scriptsize}};
  \node[state] (x2) [below = 0.5cm of x1] {\begin{scriptsize}\begin{tabular}{c}
    \textcolor{Green}{\texttt{occupancy}: \textbf{FALSE}} \\
    \texttt{camera}: \textbf{OFF} \\
    \texttt{lock}: \textbf{LOCKED} \\
    \texttt{tv}: \textbf{OFF} \\
    \texttt{fan}: \textbf{OFF} \\
    \texttt{ac}: \textbf{OFF} \\
    \texttt{temperature}: \textbf{25} \\
    \textcolor{Red}{\texttt{light1}: \textbf{OFF}} \\
    \textcolor{Red}{\texttt{light2}: \textbf{OFF}} \\
    \textcolor{Red}{\texttt{location}: \textbf{1}}
  \end{tabular}\end{scriptsize}};

  \node[state] (y0) [right = 0.5cm of x0] {\begin{scriptsize}\begin{tabular}{c}
    \textcolor{Green}{\texttt{occupancy}: \textbf{TRUE}} \\
    \texttt{camera}: \textbf{OFF} \\
    \texttt{lock}: \textbf{LOCKED} \\
    \texttt{tv}: \textbf{OFF} \\
    \texttt{fan}: \textbf{OFF} \\
    \texttt{ac}: \textbf{OFF} \\
    \texttt{temperature}: \textbf{25} \\
    \textcolor{Red}{\texttt{light1}: \textbf{OFF}} \\
    \textcolor{Red}{\texttt{light2}: \textbf{OFF}} \\
    \textcolor{Red}{\texttt{location}: \textbf{1}}
  \end{tabular}\end{scriptsize}};
  \node[state] (y1) [below = 0.5cm of y0] {\begin{scriptsize}\begin{tabular}{c}
    \textcolor{Green}{\texttt{occupancy}: \textbf{TRUE}} \\
    \texttt{camera}: \textbf{OFF} \\
    \texttt{lock}: \textbf{LOCKED} \\
    \texttt{tv}: \textbf{ON} \\
    \texttt{fan}: \textbf{OFF} \\
    \texttt{ac}: \textbf{OFF} \\
    \texttt{temperature}: \textbf{25} \\
    \textcolor{Red}{\texttt{light1}: \textbf{OFF}} \\
    \textcolor{Red}{\texttt{light2}: \textbf{OFF}} \\
    \textcolor{Red}{\texttt{location}: \textbf{1}}
  \end{tabular}\end{scriptsize}};
  \node[state] (y2) [below = 0.5cm of y1] {\begin{scriptsize}\begin{tabular}{c}
    \textcolor{Green}{\texttt{occupancy}: \textbf{TRUE}} \\
    \texttt{camera}: \textbf{OFF} \\
    \texttt{lock}: \textbf{LOCKED} \\
    \texttt{tv}: \textbf{ON} \\
    \texttt{fan}: \textbf{OFF} \\
    \texttt{ac}: \textbf{OFF} \\
    \texttt{temperature}: \textbf{25} \\
    \textcolor{Red}{\texttt{light1}: \textbf{OFF}} \\
    \textcolor{Red}{\texttt{light2}: \textbf{ON}} \\
    \textcolor{Red}{\texttt{location}: \textbf{1}}
  \end{tabular}\end{scriptsize}};

  \node (t0) [left = 0.5cm of x0] {\begin{scriptsize}$T$\end{scriptsize}};
  \node (t1) [left = 0.5cm of x1] {\begin{scriptsize}$T+1$\end{scriptsize}};
  \node (t2) [left = 0.5cm of x2] {\begin{scriptsize}$T+2$\end{scriptsize}};

  \path (x0) edge [right] node {} (x1);
  \path (x1) edge [right] node {} (x2);
  \path (y0) edge [right] node {\begin{scriptsize}R6\end{scriptsize}} (y1);
  \path (y1) edge [right] node {\begin{scriptsize}R8\end{scriptsize}} (y2);
  \end{tikzpicture}
  \caption{Example of privacy leakage. Different PRIVATE values
    (green) lead to different PUBLIC values (red) after some 
    transitions.}\vspace{-0.5cm}
  \label{fig:example-of-information-leakage}
\end{figure}

\paragraph{Privilege escalation.}
A security policy defines the expected behaviors of devices, and can
be represented by conditions that either must be or must not be
satisfied.
They are stated in the form of ``$device_1$ is
(not) $state_1$ and/or $device_2$ is (not) $state_2$ and/or ...''.
To improve usability, users only need to select the values of
$device_i$, $state_i$ and the logical connective (and, or, not); \name will then convert it into LTL or CTL.
For example,  $AG (\texttt{lock} = \textbf{LOCKED} \lor
\texttt{camera} = \textbf{ON} ) =
AG \lnot (\texttt{lock} = \textbf{UNLOCKED} \wedge
\texttt{camera} = \textbf{OFF} )$ in CTL expresses that the policy
``smart lock is unlocked but surveillance camera is off'' should
not happen.




\paragraph{Privacy leakage.}
\begin{figure}[t]
  \centering
  \begin{subfigure}[t]{0.4\linewidth}
    \begin{tikzpicture} [->,>=stealth',semithick]
    \tikzstyle{state} = [draw, thick, fill=white, rectangle];
    \node[state,fill=gray!20] (x0) {\begin{scriptsize}\begin{tabular}{c}
      \textcolor{Green}{\texttt{occupancy}: \textbf{FALSE}} \\
      \texttt{temperature}: \textbf{25} \\
    \end{tabular}\end{scriptsize}};
    \node[state] (x1) [below = 0.5cm of x0] {\begin{scriptsize}\begin{tabular}{c}
      \textcolor{Green}{\texttt{occupancy}: \textbf{TRUE}} \\
      \texttt{temperature}: \textbf{25} \\
    \end{tabular}\end{scriptsize}};

    \path (x0) edge [right] node {} (x1);
    \path (x0) edge [in=175, out=185, distance=0.6cm] node {} (x0);
    \path (x1) edge [in=175, out=185, distance=0.6cm] node {} (x1);
    \end{tikzpicture}
    \caption{Original FSM}
    \label{subfig:original-fsm}
  \end{subfigure}
  \begin{subfigure}[t]{0.4\linewidth}
    \begin{tikzpicture} [->,>=stealth',semithick]
    \tikzstyle{state} = [draw, thick, fill=white, rectangle];
    \node[state] (y0) {\begin{scriptsize}\begin{tabular}{c}
      \textcolor{Green}{\texttt{occupancy'}: \textbf{FALSE}} \\
      \texttt{temperature'}: \textbf{25} \\
    \end{tabular}\end{scriptsize}};
    \node[state,fill=gray!20] (y1) [below = 0.5cm of y0] {\begin{scriptsize}\begin{tabular}{c}
      \textcolor{Green}{\texttt{occupancy'}: \textbf{TRUE}} \\
      \texttt{temperature'}: \textbf{25} \\
    \end{tabular}\end{scriptsize}};

    \path (y0) edge [right] node {} (y1);
    \path (y0) edge [in=175, out=185, distance=0.6cm] node {} (y0);
    \path (y1) edge [in=175, out=185, distance=0.6cm] node {} (y1);
    \end{tikzpicture}
    \caption{Cloned FSM}
    \label{subfig:cloned-fsm}
  \end{subfigure}
  \\
  \vspace{0.3cm}
  \begin{subfigure}[t]{0.8\linewidth}
    \begin{tikzpicture} [->,>=stealth',semithick]
    \tikzstyle{state} = [draw, thick, fill=white, rectangle];
    \node[state] (x0y0) {\begin{scriptsize}\begin{tabular}{c}
      \textcolor{Green}{\texttt{occupancy}: \textbf{FALSE}} \\
      \texttt{temperature}: \textbf{25} \\
      \textcolor{Green}{\texttt{occupancy'}: \textbf{FALSE}} \\
      \texttt{temperature'}: \textbf{25} \\
    \end{tabular}\end{scriptsize}};
    \node[state, fill=gray!20] (x0y1) [right = 0.5cm of x0y0] {\begin{scriptsize}\begin{tabular}{c}
      \textcolor{Green}{\texttt{occupancy}: \textbf{FALSE}} \\
      \texttt{temperature}: \textbf{25} \\
      \textcolor{Green}{\texttt{occupancy'}: \textbf{TRUE}} \\
      \texttt{temperature'}: \textbf{25} \\
    \end{tabular}\end{scriptsize}};
    \node[state] (x1y0) [below = 0.5cm of x0y0] {\begin{scriptsize}\begin{tabular}{c}
      \textcolor{Green}{\texttt{occupancy}: \textbf{TRUE}} \\
      \texttt{temperature}: \textbf{25} \\
      \textcolor{Green}{\texttt{occupancy'}: \textbf{FALSE}} \\
      \texttt{temperature'}: \textbf{25} \\
    \end{tabular}\end{scriptsize}};
    \node[state] (x1y1) [right = 0.5cm of x1y0] {\begin{scriptsize}\begin{tabular}{c}
      \textcolor{Green}{\texttt{occupancy}: \textbf{TRUE}} \\
      \texttt{temperature}: \textbf{25} \\
      \textcolor{Green}{\texttt{occupancy'}: \textbf{TRUE}} \\
      \texttt{temperature'}: \textbf{25} \\
    \end{tabular}\end{scriptsize}};

    \path (x0y0) edge [right] node {} (x0y1);
    \path (x0y0) edge [right] node {} (x1y0);
    \path (x0y0) edge [right] node {} (x1y1);
    \path (x0y1) edge [right] node {} (x1y1);
    \path (x1y0) edge [right] node {} (x1y1);
    \path (x0y0) edge [in=175, out=185, distance=0.6cm] node {} (x0y0);
    \path (x1y0) edge [in=175, out=185, distance=0.6cm] node {} (x1y0);
    \path (x0y1) edge [in=5, out=355, distance=0.6cm] node {} (x0y1);
    \path (x1y1) edge [in=5, out=355, distance=0.6cm] node {} (x1y1);
    \end{tikzpicture}
    \caption{Product FSM}
    \label{subfig:product-fsm}
  \end{subfigure}
  \caption{Example of a product machine. The initial states are shaded.
    On the product FSM, two parallel traces of the original FSM can be
    viewed as a single trace, which makes our privacy policy
    expressible in LTL/CTL.}\vspace{-0.5cm}
\end{figure}
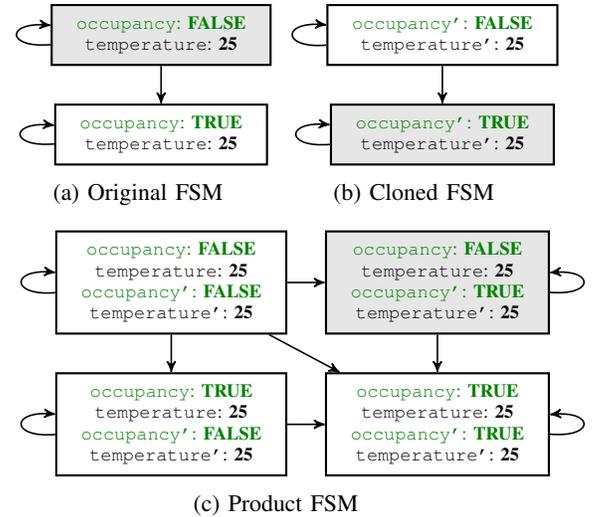
The privacy policy defines the level of privacy for each device,
which can be expressed by assigning a label
(PRIVATE, PUBLIC, or OTHER) to each attribute.

A device attribute labeled as PRIVATE indicates that the attribute
contains confidential information and should be protected from an
attacker's observation, while the PUBLIC label indicates that an
attacker may end up observing the attribute through hacking or some
local observations (e.g., lights can be observed from outside at night). We
propose the OTHER label for those attributes that are neither
confidential (PRIVATE) nor vulnerable (PUBLIC) from the user's
perspectives (i.e., user does not care about the information leakage).
Although the OTHER class complicates the analysis process and cannot
be handled by prior
methods~\cite{Gilles2004,dimitrova2012model,Milijana2017}, it allows
us to accurately model real-world settings.

In the motivating example, the attribute \texttt{occupancy}, which
indicates whether someone is home, is considered private. However,
as shown in Figure~\ref{fig:example-of-information-leakage}, attackers
can infer the actual value of this PRIVATE attribute (colored in green)
at time $T$ by observing the PUBLIC attributes (colored in red) at
time $T+2$, because different values of the PRIVATE attribute trigger
different automation rules and eventually lead to different values of
the PUBLIC attributes. That is, if the value of \texttt{light2}
is \textbf{ON}, the value of \texttt{occupancy} should
be \textbf{TRUE}. Otherwise, it should be \textbf{FALSE}.

Inspired by this observation, we want to ensure that any two
states that are only different in PRIVATE attributes in the
same environmental changes should stay indistinguishable from
the adversary's perspective at any future time. That is, the values of the
PUBLIC attributes should stay the same between these two traces at any
moment. If not, the adversary can tell the two traces apart and infer
the values of PRIVATE attributes. The environmental changes should be
equivalent in both traces because attackers can only observe the
change in real life.

Formally, \name defines security against privacy leakage as follows:
\begin{displaymath}
\begin{array}{l}
\forall s_0, s_0' \in S, t > 0 \\
\left \{
  \begin{tabular}{ll}
    $s_0 =_{PUBLIC,OTHER} s_0'$ & \\
    $s_t(a_i) = s_t'(a_i)$ & $\forall a_i \in A_R$ \\
  \end{tabular}
  \right. \\
\implies s_t =_{PUBLIC} s_t',
\end{array}
\end{displaymath}
where $=_{PUBLIC}$ and $=_{PUBLIC,OTHER}$ stand for the equivalence over PUBLIC only
and PUBLIC and OTHER attributes,
respectively. 

Hence, to check whether a PRIVATE attribute can be leaked, we can modify
the PRIVATE attribute and see if the trace is altered from the attacker's perspective. Nonetheless, these kinds of properties cannot be expressed in
ordinary LTL or CTL because it requires pairwise comparison between
two traces of the finite state machine. 

To overcome this challenge, we construct a product machine as follows.
We first build an almost equivalent copy of the original FSM, and
their differences reside in the PRIVATE values of the initial
states.
For example, suppose the original FSM is Figure~\ref{subfig:original-fsm},
then the copy one named CLONE would be Figure~\ref{subfig:cloned-fsm}. (Shaded
states are the initial states.)
Formally, $CLONE \equiv (S, \longrightarrow, I')$, and $I'
= \{s' | s' =_{L,O} s, \forall s' \in S, s \in I\}$.
Then, the states of product machine $PM$ are Cartesian products of the
states of the two state machines, as in Figure~\ref{subfig:product-fsm}.
The initial state is the state that consists of the initial states of
original and cloned FSM. The transitions can be seen as the union
of the transitions from both machines. That is,
$PM \equiv (S \times S, \longrightarrow',
I \times I')$, where $\longrightarrow' = \{ ((s_i, s_j), (s_i', s_j')) | (s_i, s_j) \in
\longrightarrow \wedge (s_i', s_j') \in \longrightarrow \}$.


We also add constraints
to ensure that the original and cloned machines undergo the same
environmental changes. To this end, the problem is reduced to whether
the product machine can arrive at a state where its two internal
states are different in the PUBLIC attributes, which is a reachability
problem and can be expressed in ordinary LTL or CTL.
Although building a product machine increases the state
space\footnote{If the original state machine is $O(2^N)$, then the
product machine will become $O(2^N) * O(2^N) = O(4^N)$, where $N$ is
the number of attributes.}, our evaluation shows that the verification
can still be efficiently performed with our optimization techniques.

Our product FSM construction is similar to the self-composition
approach proposed by Gilles et al.~\cite{Gilles2004}. However, it is
unclear whether their program-execution-based non-interference
definition can model the event-driven scenarios in trigger-action
programming. In addition, \name avoids the additional cost of
transforming a program to a state machine because the security
properties of trigger-action programming can be directly defined over
a transition system instead of a program.

\subsection{Optimization}
\label{ssec:optimization}
To enable timely verification, \name's optimization component aims to
simplify the input (a model of a smart space) by removing redundant
attribute values, rules, and devices. As the input to the model
checker is reduced, the verification process can be accelerated.
We explain in detail how \name reduces the size of the problem
by \emph{grouping} attribute values that always cause the same effects,
and by \emph{pruning} devices and rules irrelevant to the security
policy.

Generally speaking, our practice shares the same spirit with
common techniques to address state explosion (e.g. grouping equivalences).
However, thanks to the simple structure of trigger-action programs,
we can effectively address the state explosion problem in model checking
using static code analysis. We first obtain  high-level program
semantics using static analysis, and then efficiently construct an
abstraction of the state-transition system without spurious
counterexamples; thus, no further refinement is
needed~\cite{Clarke2000}.  In addition, as our application scenarios
consider non-technical users who specify their own rules and security
policies, we further consider optimizations for simple security
polices (in contrast to complex ones with temporal qualifiers).




\begin{table}[t]
  \centering
    \caption{Grouping of variables. Given a set of rules, values of variables are grouped
      together if they trigger same actions, 
      reducing the number of states to be considered in FSM.}
  \resizebox{\linewidth}{!}{%
  \begin{scriptsize}
  \begin{threeparttable}
    \begin{tabular}{|c|c|c|c|}
      \hline
      \textbf{Attributes} & \textbf{Possible values} & \textbf{Constraints} & \textbf{Grouped values} \\
      \hline \hline
      \multirow{2}{*}{\texttt{light1}} & \multirow{2}{*}{\textbf{ON}, \textbf{OFF}} & $\texttt{light1} = \textbf{OFF}$ & \multirow{2}{*}{\textbf{ON}, \textbf{OFF}} \\
      & & $\texttt{light1} \leftarrow \textbf{OFF}$ & \\
      \hline
      \texttt{camera} & \textbf{ON}, \textbf{OFF} & $\texttt{camera} = \textbf{ON}$ & \textbf{ON}, \textbf{OFF} \\
      \hline
      \multirow{2}{*}{\texttt{location}} & \multirow{2}{*}{$\textbf{0},...,\textbf{10}$\tnote{*}} & $\texttt{location} = \textbf{0}$ & \multirow{2}{*}{\textbf{0},\textbf{OTHERS}} \\
      & & $\texttt{location} \neq \textbf{0}$ & \\
      \hline
      \multirow{2}{*}{\texttt{lock}} & \textbf{LOCKED} & \multirow{2}{*}{$\texttt{lock} = \textbf{LOCKED}$} & \textbf{LOCKED} \\
      & \textbf{UNLOCKED} & & \textbf{UNLOCKED} \\
      \hline
      \multirow{2}{*}{\texttt{occupancy}} & \multirow{2}{*}{\textbf{TRUE}, \textbf{FALSE}} & $\texttt{occupancy} = \textbf{TRUE}$ & \multirow{2}{*}{\textbf{TRUE}, \textbf{FALSE}} \\
      & & $\texttt{occupancy} = \textbf{FALSE}$ & \\
      \hline
      \multirow{2}{*}{\texttt{tv}} & \multirow{2}{*}{\textbf{ON}, \textbf{OFF}} & $\texttt{tv} = \textbf{ON}$ & \multirow{2}{*}{\textbf{ON}, \textbf{OFF}} \\
      & & $\texttt{tv} = \textbf{OFF}$ & \\
      \hline
      \multirow{2}{*}{\texttt{light2}} & \multirow{2}{*}{\textbf{ON}, \textbf{OFF}} & $\texttt{light2} \leftarrow \textbf{ON}$ & \multirow{2}{*}{\textbf{ON}, \textbf{OFF}} \\
      & & $\texttt{light2} \leftarrow \textbf{OFF}$ & \\
      \hline
      \texttt{fan} & \textbf{ON}, \textbf{OFF} & & \textbf{ALL} \\
      \hline
      \texttt{ac} & \textbf{ON}, \textbf{OFF} & & \textbf{ALL} \\
      \hline
      \multirow{3}{*}{\texttt{temperature}} & \multirow{3}{*}{$\textbf{23},...,\textbf{33}$\tnote{*}} & $\texttt{temperature} \geq \textbf{28}$ & \textbf{26..27},\textbf{28..31} \\
      & & $\texttt{temperature} \geq \textbf{32}$ & \textbf{32..33} \\
      & & $\texttt{temperature} \leq \textbf{25}$ & \textbf{23..25} \\
      \hline
    \end{tabular}
    \begin{tablenotes}
      \item[*] Short-lived windows for \texttt{location} and \texttt{temperature} are $[\textbf{0},\textbf{10}]$ and $[\textbf{23},\textbf{33}]$ respectively.
    \end{tablenotes}
  \end{threeparttable}
  \end{scriptsize}}

  \label{tab:grouping-of-variables}
\end{table} 

\paragraph{Grouping.}
In trigger-action programming, two attribute values can be
considered \emph{equivalent} if they always trigger the same actions,
regardless of the other attributes. Hence, we can reduce the number of
states needed to be considered by grouping equivalent values into
subgroups (or meta-values) and then rewrite the automation rules
using the new meta-values. Algorithm~\ref{alg:grouping} shows the
pseudo-code. First, we collect all constraints from the rules (Lines
2-9) and the policy (Lines 10-13). Then, the acquired constraints are
sorted for each attribute, which is used to classify possible
values into meta-values (Lines 14-18).  Here, we build a map to convert
between meta-values and original values. Finally, we translate rules
to equivalencies by looking up the map (Lines 19-24). Overall,
sorting the acquired constraints for each attribute takes most
of the time, and the number of possible constraints is proportional to
the number of rules. For a single attribute, the maximum number of
constraints is $O(kM)$, where $k$ is the maximum number of constraints
in a rule and $M$ is the number of rules. Hence, the time complexity
is $O(kMN\log kM)$, where $N$ is the number of attributes.

Table~\ref{tab:grouping-of-variables} and
Table~\ref{tab:rewriting-of-rules} show the meta-values and rewritten
rules after the grouping method is applied to
the example in \S\ref{ssec:a-motivating-example},
respectively. The temperature values $\leq 25$ are considered
equivalent because they all trigger rule R12. The possible values of
the attribute \texttt{temperature} are divided into
four subgroups, thereby the number of states is reduced to four.  In
rule R10, the trigger $\texttt{temperature} \geq \textbf{28}$ is converted
to $\texttt{temperature} = \textbf{28..31} \vee \texttt{temperature}
= \textbf{32..33}$ and action $\texttt{fan} \leftarrow \textbf{ON}$ is
the same as $\texttt{fan} \leftarrow \textbf{ALL}$ because its value
does not affect the execution of any rules.

Note that constraints in policies should also be taken into
consideration as they will also affect the grouping results.





\begin{table}[t]
  \centering
  \caption{Rule rewriting. Rules in Table~\ref{tab:example-rules}
    are rewritten w.r.t. the grouped values in Table~\ref{tab:grouping-of-variables}.}
  \begin{scriptsize}
  \begin{tabular}{|c|c|c|}
    \hline
    \textbf{Rule} & \textbf{Trigger} & \textbf{Action} \\
    \hline \hline
    \multirow{3}{*}{R10} & Temperature is a little high & Turn on the fan \\
    & $\texttt{temperature} = \textbf{28..31}$ & \multirow{2}{*}{$\texttt{fan} \leftarrow \textbf{ALL}$} \\
    & $\vee \texttt{temperature} = \textbf{32..33}$ & \\
    \hline
    \multirow{2}{*}{R11} & Temperature is high & Turn on the air conditioner \\
    & $\texttt{temperature} = \textbf{32..33}$ & $\texttt{ac} \leftarrow \textbf{ALL}$ \\
    \hline
    \multirow{3}{*}{R12} & \multirow{2}{*}{Temperature is low} & Turn off the fan \\
    & & Turn off the air conditioner \\
    & $\texttt{temperature} = \textbf{23..25}$ & $\texttt{fan} \leftarrow \textbf{ALL}, \texttt{ac} \leftarrow \textbf{ALL}$ \\
    \hline
  \end{tabular}
  \end{scriptsize}

  \label{tab:rewriting-of-rules}
\end{table}  

\begin{algorithm}[t]
  \scriptsize
  \setstretch{1}
  \KwData{\\
    $A$ is a list of attributes\\
    $R$ is a list of rules\\
    $P$ is the policy to be verified}
  \KwResult{$A^{Grouped}$, $R^{Grouped}$}
  \Begin{
    $constraints = Dictionary(List)$\;
    \For{$r_i \in R$} {
      $trigger \leftarrow GetTrigger(r_i)$\;
      \For{$constraint \in trigger$} {
        $a_i \leftarrow GetTarget(constraint)$\;
        $constraints[a_i].append(constraint)$\;
      }
    }
    \For{$constraint \in P$} {
      $a_i \leftarrow GetTarget(constraint)$\;
      $constraints[a_i].append(constraint)$\;
    }
    \For{$a_i \in A$} {
      $values \leftarrow Sorted(constraints[a_i])$\;
      $a_i^{Grouped} \leftarrow GroupAttribute(a_i, values)$\;
      $A^{Grouped}.append(a_i^{Grouped})$\;
    }
    \For{$r_i \in R$}{
      $bool_i \leftarrow GetEqualTrigger(A^{Grouped}, r_i)$\;
      $assign_i \leftarrow GetEqualAction(A^{Grouped}, r_i)$\;
      $r_i^{Grouped} \leftarrow MakeRule(bool_i, assign_i)$\;
      $R^{Grouped}.append(r_i^{Grouped})$\;
    }
  }
  \caption{Grouping}
  \label{alg:grouping}
\end{algorithm}

\paragraph{Pruning.}
While grouping reduces the number of possible values of an attribute,
pruning aims to reduce the number of attributes
needed to be considered because not every device can influence the
state of another device. For instance, to check this CTL specification
$AG \lnot (lock = \textbf{UNLOCKED} \wedge camera = \textbf{OFF})$ in
the example in \S\ref{ssec:a-motivating-example}, we only need to consider devices in the front door
because all the devices inside the house have no impact (via the
automation rule) on the smart lock and the surveillance camera.


Algorithm~\ref{alg:pruning} shows the pseudo-code for pruning. To prune
irrelevant devices, we represent the relationship between devices by
building a dependency graph (Lines 2-14), in which each vertex represents an attribute,
and a directed edge from one vertex to another represents one
attribute being able to affect the other. After building a dependency graph, we
back-trace from interesting attributes (that are involved in the security
policy) to find relevant devices (Lines 15-26). Building the
dependency graph takes most of the time and requires adding edges
between trigger attributes and action attributes. The worst-case scenario is
that both the trigger and action contain all attributes for each rule.
Thus, the time complexity is $O(MN^2)$, where $M$ and $N$ are the
number of rules and attributes, respectively.

\begin{figure}[t]
  \centering
  \scriptsize
  \begin{tikzpicture} [->,>=stealth',semithick, state/.style={circle, draw, minimum size=1cm, inner sep=0pt, text width=1cm, align=center}]
  \tikzstyle{target} = [circle, draw, thick, fill=black, text=white, minimum size=1cm, inner sep=0pt, text width=1cm, align=center];
  \tikzstyle{related} = [circle, draw, thick, fill=lightgray, minimum size=1cm, inner sep=0pt, text width=1cm, align=center];
  \node[related] (light1) {\textbf{\texttt{light1}}};
  \node[target] (camera) [right = of light1] {\textbf{\texttt{camera}}};
  \node[related] (location) [below = of light1] {\textbf{\texttt{loc\\ation}}};
  \node[target] (lock) [below = of camera] {\textbf{\texttt{lock}}};
  \node[state] (occupancy) [right = 0.4cm of camera] {\texttt{occup\\ancy}};
  \node[state] (tv) [below = 0.7cm of occupancy] {\texttt{tv}};
  \node[state] (fan) [below left = 0.4cm of location] {\texttt{fan}};
  \node[state] (temperature) [right = of fan] {\texttt{tempe\\rature}};
  \node[state] (ac) [right = of temperature] {\texttt{ac}};
  \node[state] (light2) [right = 0.4cm of ac] {\texttt{light2}};

  \path (location) edge [left] node {\textbf{R1}} (light1);
  \path (location) edge [below left] node {\textbf{R1}} (camera);
  \path (location) edge [below] node {\textbf{R2}} (lock);
  \path (camera) edge [right] node {\textbf{R3}} (lock);
  \path (lock) edge [below right] node {\textbf{R4}} (light1);
  \path (light1) edge [above] node {\textbf{R5}} (camera);
  \path (occupancy) edge node {R6 R7} (tv);
  \path (tv) edge node {R8 R9} (light2);
  \path (temperature) edge node [text width=0.6cm, align=center] {R10\\R12} (fan);
  \path (temperature) edge node [text width=0.6cm, align=center] {R11\\R12} (ac);
  \end{tikzpicture}
  \caption{Dependency graph for privilege escalation. The vertices
  represent attributes and the edges represent rules. Only attributes
  which involved in the security policy (in black) and their
  dependencies (in grey) need to be considered in the verification
  process.}\vspace{-0.2cm}
  \label{fig:dependency-graph}
\end{figure}
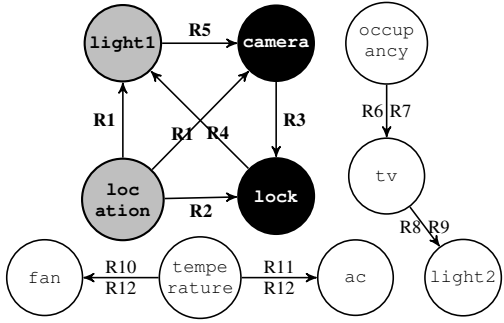

Figure~\ref{fig:dependency-graph} illustrates the dependency graph of
the example in \S\ref{ssec:a-motivating-example}. For each rule, we check the dependency between the
involved devices. For example, in rule R1, the
attributes \texttt{light1} and \texttt{camera} are affected by the
attribute \texttt{location}.
Thus, on the dependency graph, we add two directed edges
from \texttt{location} to \texttt{light1} and \texttt{camera}
respectively, and label these two edges with rule R1 to track their
relationship. During backtracking, we start from the interesting
attributes (e.g., \texttt{lock} and \texttt{camera} in the policy
$AG \lnot (\texttt{lock} = \textbf{UNLOCKED} \wedge \texttt{camera}
= \textbf{OFF} )$) and find all connected vertices (i.e., relevant
attributes) and connected edges (i.e., relevant rules).


In privacy leakage, the vulnerable devices
observable by attackers are interesting attributes, such as \texttt{light1}, \texttt{light2}
and \texttt{location} in the example in \S\ref{ssec:a-motivating-example}. We further accelerate the
verification of privacy leakage by filtering out \LOW attributes that
are unreachable from any \HIGH attribute on the dependency graph,
because there is no leak if the \HIGH attributes have no impact on the
\LOW attributes. If any \LOW attribute can be reached, we back-trace from
those reachable \LOW vertices. Hence, we can perform the verification
with only three attributes after back-tracing from \texttt{light2} in
the example in \S\ref{ssec:a-motivating-example}.

\begin{algorithm}[t]
  \scriptsize
  \setstretch{1}
  \KwData{\\
    $A$ is a list of attributes\\
    $R$ is a list of rules\\
    $P$ is the policy to be verified}
  \KwResult{$A^{Related}$, $R^{Related}$}
  \Begin{
    $graph \leftarrow Graph()$\;
    \For{$a_i \in A$}{
        $graph.addNode(a_i)$\;
    }
    \For{$r_i \in R$}{
      $bool\_attributes \leftarrow GetTriggerAttributes(r_i)$\;
      $assign\_attributes \leftarrow GetActionAttributes(r_i)$\;
      \For{$a_j \in bool\_attributes$}{
        \For{$a_k \in assign\_attributes$}{
          $graph.addEdge(a_j, a_k, r_i)$\;
        }
      }
    }
    $unexplored\_nodes \leftarrow GetAssociatedAttributes(Policy)$\;
    \While{$unexplored\_nodes \neq \emptyset$}{
      $a \leftarrow unexplored\_nodes.pop()$\;
      $A^{Related}.add(a)$\;
      \For{$neighbor \in graph.neighbors(a)$}{
        \If{$neighbor \notin A^{Related}$}{
          $unexplored\_nodes.add(neighbor)$\;
        }
        $r \leftarrow graph.getEdgeMark(a, neighbor)$\;
        $R^{Related}.add(r)$\;
      }
    }
  }
  \caption{Pruning}
  \label{alg:pruning}
\end{algorithm}

\subsection{Mitigation}
\label{ssec:mitigation}


Ideally, if one could fix every vulnerable device immediately, the
attacker would be unable to increase the attack surface by exploiting
automation rules. However, patching in a timely manner is
challenging, and devices may also have zero-day vulnerabilities. Hence,
in this work, we discuss mitigations that can be achieved by updating
automation rules, which include removing or modifying exploitable
ones with users' consent. The question then becomes which to remove
or how to modify them.



A straightforward approach is to put every rule involved in the
identified attacks into a watchlist, and request confirmation from the user whenever
a watchlisted rule is about to be executed. Although this can indeed
prevent attacks, it undermines the convenience of this
system.


To reduce the level of inconvenience, we would like to add as few
rules to the watchlist as possible.  For each attack chain, it is
sufficient to stop the attack by blocking at least one rule in the
chain. Also, one rule can appear in multiple attack chains. Therefore,
one possible approach is to first determine all the attack chains,
then apply some optimization techniques or greedy strategies to
minimize the blocked rules.

Nonetheless, the question remains of how to effectively figure out all the
attack traces, because model-checking tools like NuSMV report only one
counterexample at a time. It is also unclear what the selection
algorithm would be like. We will leave these problems for future research.
On the other hand, Salus~\cite{Liang2016} proposed a solution that
first parameterizes trigger conditions, and then finds a feasible
configuration using model checking. However, such an approach also
suffers from usability problems, because Salus needs user permission to
accept or reject each set of feasible parameters one by one, and there
may be multiple sets of feasible parameters.




\section{Evaluation and Implementation}
\label{sec:experiment}
We evaluate \name's security and performance using analytical
comparisons (\S\ref{ssec:security-comparison})
and experiments
(\S\ref{ssec:performance-evaluation}-\ref{ssec:accuracy-evaluation}) on a prototype implementation
described in \S\ref{ssec:implementation}.


\subsection{Security comparison}
\label{ssec:security-comparison}

\begin{table}[t]
  \newcommand{\specialcell}[2][c]{\begin{tabular}[#1]{@{}c@{}}#2\end{tabular}}
  \centering
  \caption{Comparison between \name and related work. O: supported; X: not supported; $\triangle$: inaccurate; $\diamondsuit$: can be extended.}  
  \resizebox{\linewidth}{!}{%
  \begin{scriptsize}
  \begin{threeparttable}
    \begin{tabular}{|l|c|c|c|c|}
      \hline
      \textbf{Properties} & \textbf{\name} & \specialcell{\textbf{Milijana}\\et al.~\cite{Milijana2017}} & \textbf{SIFT}~\cite{Liang2015} & \textbf{Salus}~\cite{Liang2016} \\
      \hline \hline
      Consider attacks & O & O & X\tnote{*} & X\tnote{*} \\
      \hline
      Detect privacy leakage & O & $\triangle$ & X & X \\
      \hline
      Detect privilege escalation & O & X & $\diamondsuit$ & $\diamondsuit$ \\
      \hline
      Detect integrity attack & $\diamondsuit$ & $\triangle$ & X & X \\
      \hline
      Consider device states & O & X & O & O\\
      \hline
      Analysis technique & \specialcell{Model\\checking} & \specialcell{Information\\flow} & \specialcell{Symbolic\\execution} & \specialcell{Model\\checking} \\
      \hline
      Verification Speed & Fast & Very fast & Slow & Slow \\
      \hline
    \end{tabular}
    \begin{tablenotes}
      \item[*] focuses on the reliability problem (i.e. policy validation or conflict detection) in trigger-action programming.
    \end{tablenotes}
  \end{threeparttable}
  \end{scriptsize}}

  \label{tab:comparison-between-veriiot-and-related-works}
\end{table}

We compare \name with closely related work~\cite{Milijana2017,
  Liang2016, Liang2015} and the results are summarized in
Table~\ref{tab:comparison-between-veriiot-and-related-works}.  The
integrity attack is the opposite of the privacy leakage attack;
devices with attributes labeled \HIGH
should not be affected by vulnerable devices.

To identify potential integrity and privacy violations in automation
rules, Milijana et al.~\cite{Milijana2017} define information flow
lattices and check if information can flow from a more restricted
trigger to a less restricted action. However, as their work focuses on
static analysis on automation rules and does not consider the actual
attribute values, their method cannot detect whether the system will
enter an unauthorized state in the future (privilege
escalation).
Also, static analysis may produce false positives when detecting
privacy leakage, since the actual values of devices are omitted.

As shown in Table~\ref{tab:rules-for-case-study}, R3 will raise a
false alarm because the information is propagated from restricted to
public, when in reality, it will never be triggered since the volume
will always be higher than 0. Another example is R4. Whether it will
leak information depends on the parameters. If the second condition is
set to $\texttt{volume} \leq \textbf{100}$, the attacker cannot
derive information about the volume value by monitoring the status of
the LED. Hence, the location of this user will not be disclosed.
We label both actions as public, because information can be leaked
  not only from publicly-observable devices but from vulnerable
  devices, whose information can be directly accessed by the
  attacker. For example, some smart bands have been reported with
  vulnerabilities~\cite{Classen2018}. Once
  attackers break into a device, its information is considered
  public.
 
\begin{table}[t]
  \centering
  \caption{Rules for case study. R3 will raise a false alarm in
  static analysis because the volume is always greater than or equal to 50 at
  runtime and will never trigger R3.}
  \resizebox{\linewidth}{!}{%
  \begin{scriptsize}
    \setlength\tabcolsep{3pt}
    \begin{tabular}{|c|c|c|}
      \hline
      \textbf{Rule} & \textbf{Trigger} & \textbf{Action} \\
      \hline \hline
      \multirow{2}{*}{R1} & User is at home (\emph{private}) & Set the volume (\emph{restricted}) \\
      & $\texttt{location} = \textbf{HOME}$ & $\texttt{volume} \leftarrow \textbf{50}$ \\
      \hline
      \multirow{2}{*}{R2} & User is outside (\emph{private}) & Set the volume (\emph{restricted})\\
      & $\texttt{location} \neq \textbf{HOME}$ & $\texttt{volume} \leftarrow \textbf{100}$ \\
      \hline
      \multirow{2}{*}{R3} & Any phone call missed with low volume (\emph{restricted}) & Set band to vibrate (\emph{public}) \\
      & $\texttt{missed\_call} = \textbf{TRUE} \wedge \texttt{volume} \leq \textbf{0}$ & $\texttt{band\_vibration} \leftarrow \textbf{TRUE}$ \\
      \hline
      \multirow{2}{*}{R4} & Any phone call missed with low volume (\emph{restricted}) & Blink the band LED (\emph{public}) \\
      & $\texttt{missed\_call} = \textbf{TRUE} \wedge \texttt{volume} \leq \textbf{50}$ & $\texttt{band\_led} \leftarrow \textbf{BLINK}$ \\
      \hline
    \end{tabular}
  \end{scriptsize}
  }

  \label{tab:rules-for-case-study}
\end{table}
 
SIFT~\cite{Liang2015} and Salus~\cite{Liang2016} are designed to help
users debug trigger-action rules by verifying whether the devices'
interaction (through automation rules) meets users'
expectations. Since they focus on reliability rather than security,
their techniques cannot be directly applied to detect attacks
exploiting automation, but may be extended to check privilege
escalation to some extent.
%
%
Using symbolic execution for automated analysis, SIFT~\cite{Liang2015}
starts by transforming automation rules into IF statements in C\#, and
wrap all the rules in a while loop. Each policy specified by the users
is then encoded as an assertion and will be checked by Pex, an
automated whitebox testing tool for .NET. However, they only unroll
the while loop for a fixed number of steps ($k$), and thus may have false
negatives for violations occurring after $k$ steps.
The subsequent work, Salus~\cite{Liang2016}, adopts model-checking
techniques as we do.
However, it is unclear whether Salus can handle the growing complexity
in IoT, since their experiments show that the time needed for
verification increases exponentially after irrelevant devices are
installed.
Thus, with respect to performance, Salus can be seen as a baseline
approach without optimizations.  

On the contrary, \name is designed to detect privilege escalation and
privacy leakage. It can be easily extended to detect integrity
attacks by observing whether the values of the \HIGH attributes change
when the values of the \LOW attributes are altered. In addition, \name
takes advantage of formal model checking, which provides stronger
guarantees of reducing false negatives, and supports rule-aware
optimizations to accelerate verification.


\begin{figure*}
  \centering
  \begin{subfigure}[t]{0.25\linewidth}
    \centering
    \includegraphics[width=\linewidth]{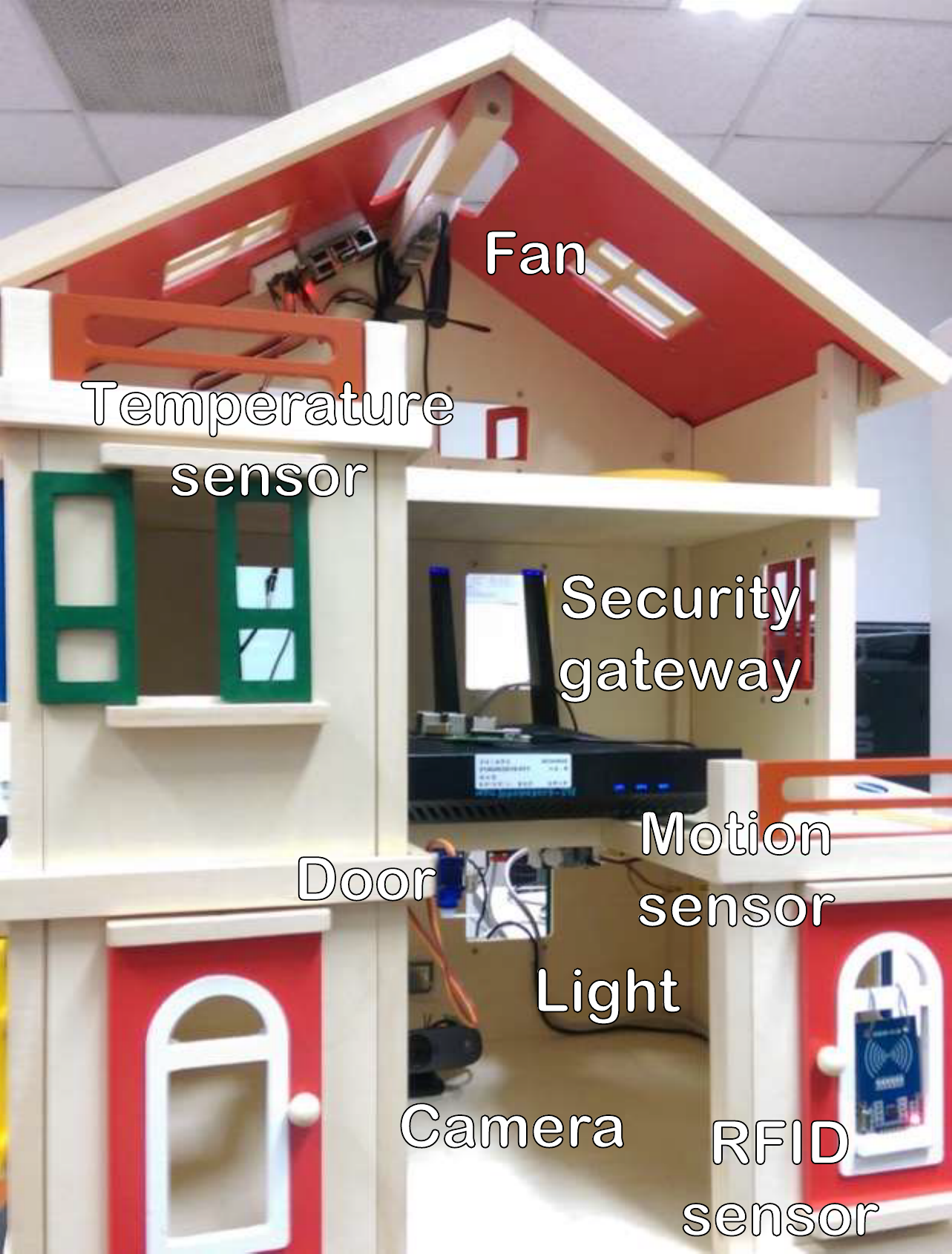}
    \caption{A toy house}
    \label{fig:implementation-of-veriiot}
  \end{subfigure}
  \begin{subfigure}[t]{0.2\linewidth}
    \centering
    \includegraphics[width=\linewidth]{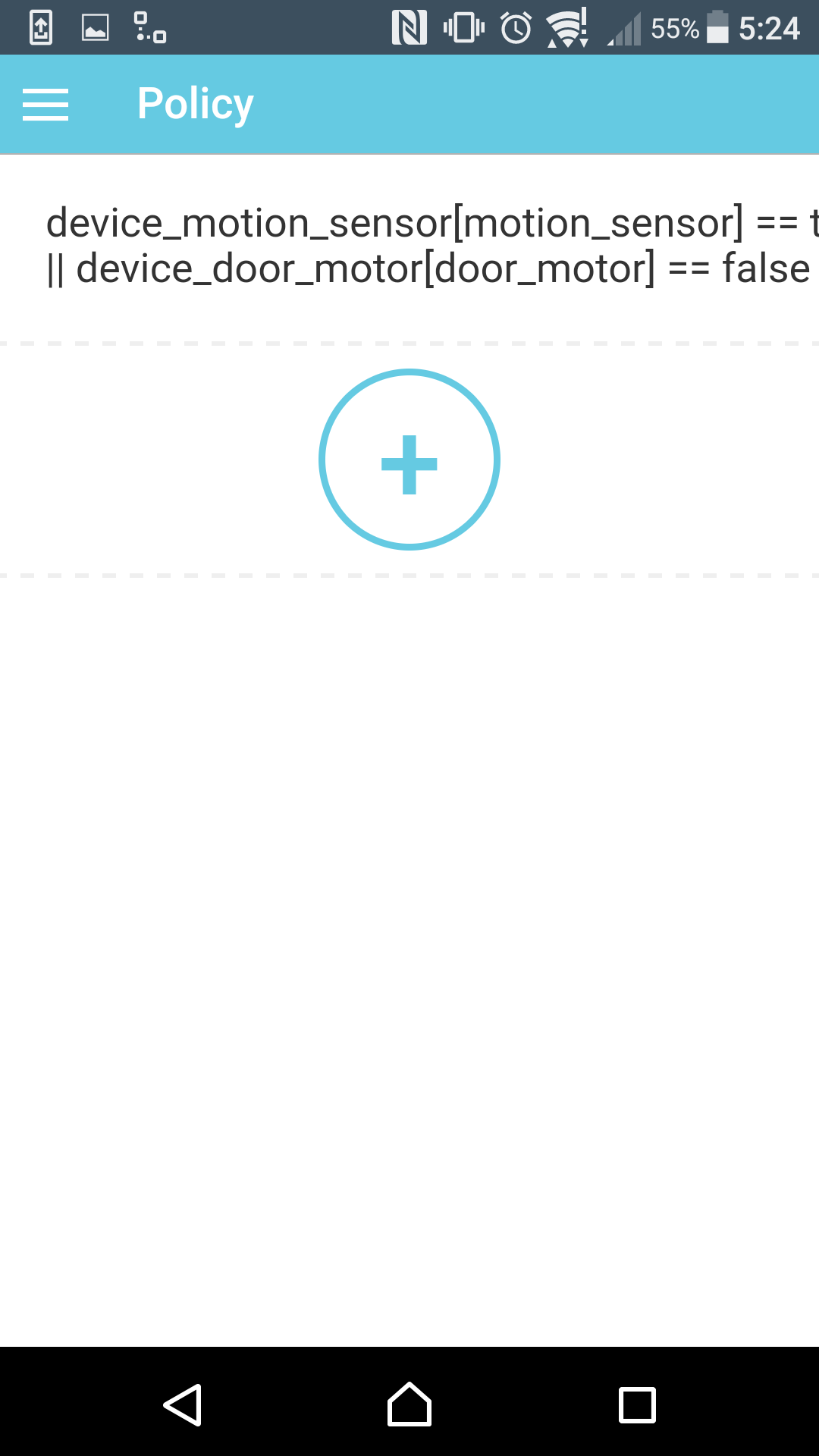}
    \caption{Security policies}
    \label{fig:security-policies-of-app}
  \end{subfigure}
  \begin{subfigure}[t]{0.2\linewidth}
    \centering
    \includegraphics[width=\linewidth]{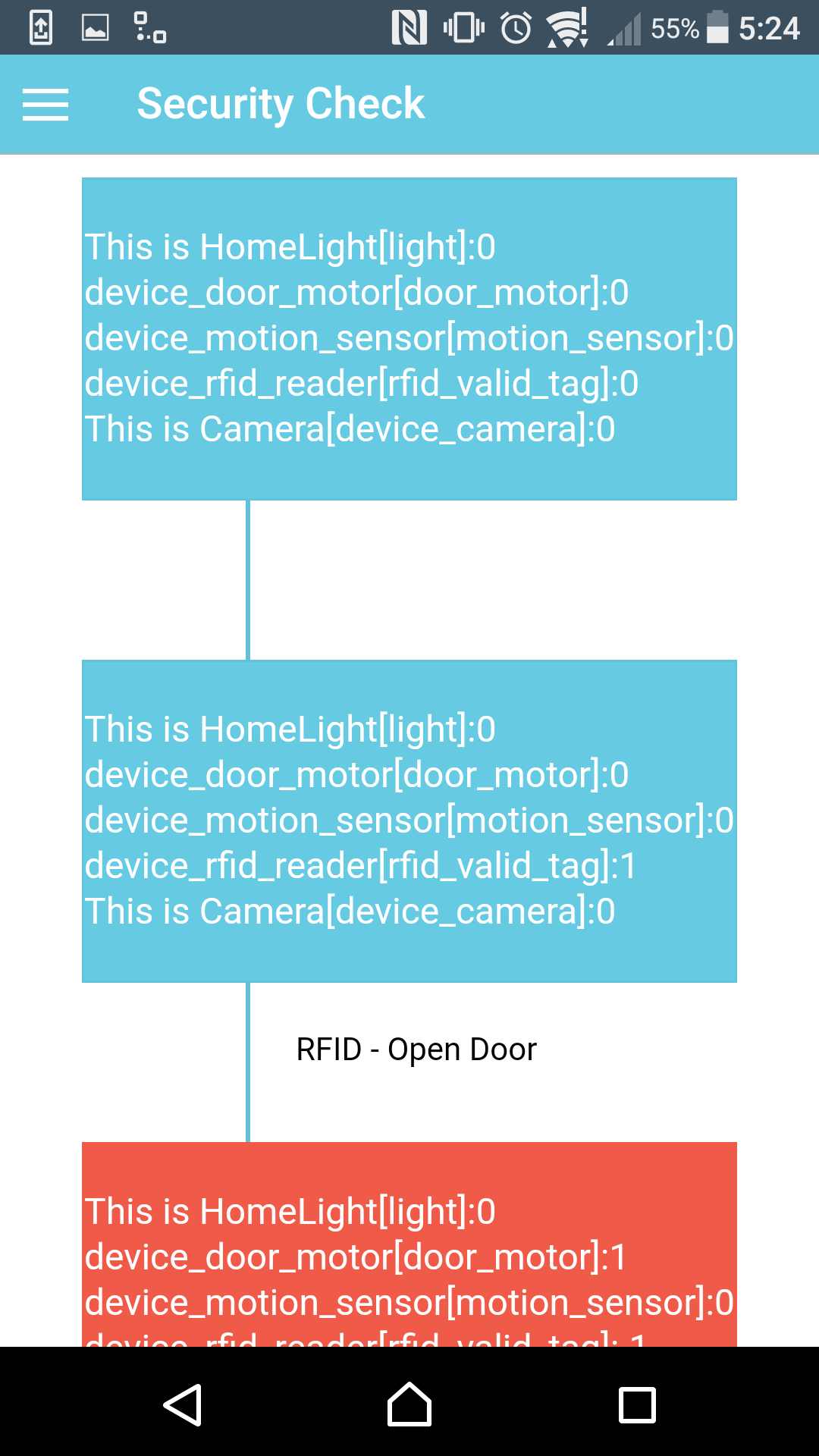}
    \caption{Check result}
    \label{fig:verification-results-of-app}
  \end{subfigure}
  \begin{subfigure}[t]{0.2\linewidth}
    \centering
    \includegraphics[width=\linewidth]{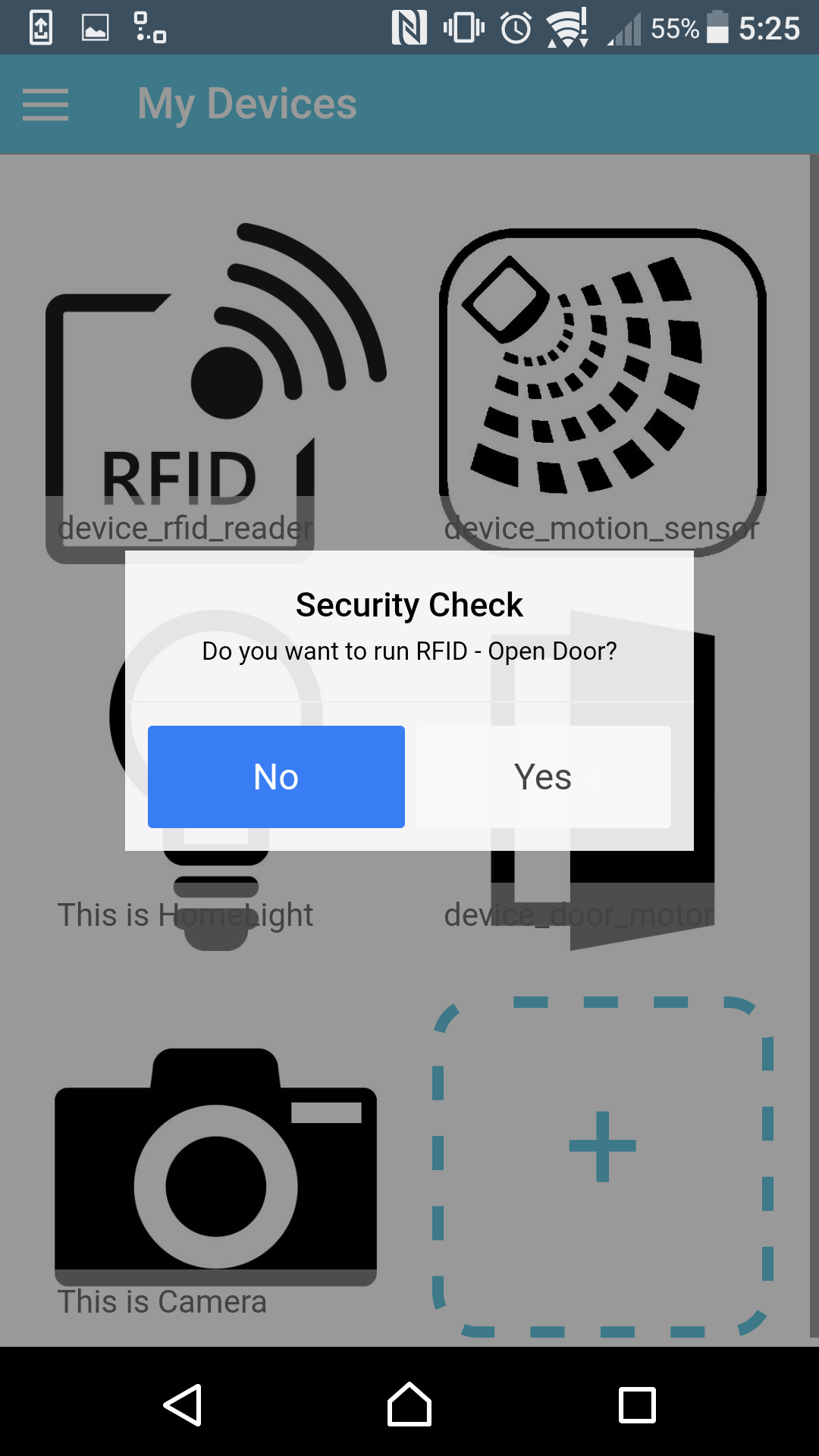}
    \caption{Security alarms}
    \label{fig:security-alarms-of-app}
  \end{subfigure}
  \caption{Proof-of-concept implementation}
  \label{fig:android-application-of-veriiot}
\end{figure*}

\subsection{Implementation}
\label{ssec:implementation}


To demonstrate the practicality of our lightweight system, we built a
smart home testbed and an Android application as shown in
Figure~\ref{fig:android-application-of-veriiot}. The major
components, an IoT gateway and \name, reside in a Raspberry Pi 2B
board.

We implement an IoT gateway to provide similar functionalities as
existing service providers. Our system is built on top of the Kura
framework~\cite{kura}, a Java/OSGi-based platform for
building IoT gateways. We add several customized control messages to
enable communication between devices and protocols like MQTT and
CoAP. To simulate a smart space, we implement several IoT devices,
such as fan, camera, temperature sensor and smart door using Arduino
Yun, Raspberry Pi and Banana Pi boards. After
devices are connected to the IoT gateway, they can share messages with
each other.

Users can utilize our Android application to monitor, control and manage
these devices. Similar to existing home automation services,
trigger-action rules can also be added using Boolean function and
assignments to enable automation. Our system will query devices'
status regularly, apply satisfied automation rules and update their
status in the application. To check which automation rules should
be applied, we use the Javaluator library~\cite{javaluator}, which is
a powerful infix expression evaluator for Java.

As discussed in {\S\ref{ssec:performance-evaluation}}, the
  security policies shown in
  Figure~{\ref{fig:security-policies-of-app}} can be set up by
  experts, modified by users, or selected from a public database, and
  {\name} will convert them into LTL or CTL for verification. The
results are shown in our application and alarms will be popped out to
request user confirmations when vulnerable automation rules are going
to be executed, as shown in
Figures~\ref{fig:verification-results-of-app}
and~\ref{fig:security-alarms-of-app}.

\paragraph{User interaction model.}
We describe how users interact with our system in a step-by-step
manner. We assume a standalone app that pulls information about
devices and automation rules from an existing automation platform
(e.g., IFTTT). A similar procedure can be described when our system
is integrated with an existing automation platform.

\begin{enumerate}
  \item A user installs our app, which comes with default
    security policies maintained by experts. The user can also modify
    or write customized policies, as shown in
    Fig.{~\ref{fig:security-policies-of-app}}.
  \item Our app runs in the background and periodically retrieves
    the most recent set of automation rules from the linked automation
    platform, and performs security checks according the specified
    security policies.
  \item When the app detects a violated security policy, a pop-up
    warning is displayed, recommending the user to review the rule
    set. When the user clicks the warning or opens our app, the user
    can see detailed information about the attack trace of the
    violation, as shown in
    Fig.{~\ref{fig:verification-results-of-app}}.
  \item If the vulnerable rule is triggered before the user
    fixes it, an alert will be displayed to request the user's explicit
    consent to execute the action, as shown in
    Fig.{~\ref{fig:security-alarms-of-app}}. This feature may
    require permissions to modify the data stored in the automation
    platform, such as temporarily disabling the action of the vulnerable rule.
\end{enumerate}

\subsection{Performance Evaluation}
\label{ssec:performance-evaluation}

Based on the implementation, we conducted a large-scale performance
evaluation using a real dataset as described below. In each run of the experiment, we
randomly sampled $N$ out of the 4,161 rules we encoded and generate security polices
with respect to these sampled rules to simulate different scenarios. \name can work with most model checkers; to
compare with previous work~\cite{Liang2016}, we use the open-sourced
model checker, NuSMV, as our backend tool.

\paragraph{Dataset.}
We selected 42 IoT-related channels from a real dataset containing 313
channels and 295,155 IFTTT rules~\cite{Ur2016}. We manually modeled
these selected channels and obtained 190 attributes and 4,161 rules.
The selected channels are listed in
Appendix~\ref{sec:channel-summary} and the encoded data are available online~\cite{our-dataset}. The total state space is roughly
$2^{650}$.

\begin{figure*}
  \centering
  \begin{subfigure}[t]{0.3\textwidth}
    \centering
    \includegraphics[width=\linewidth]{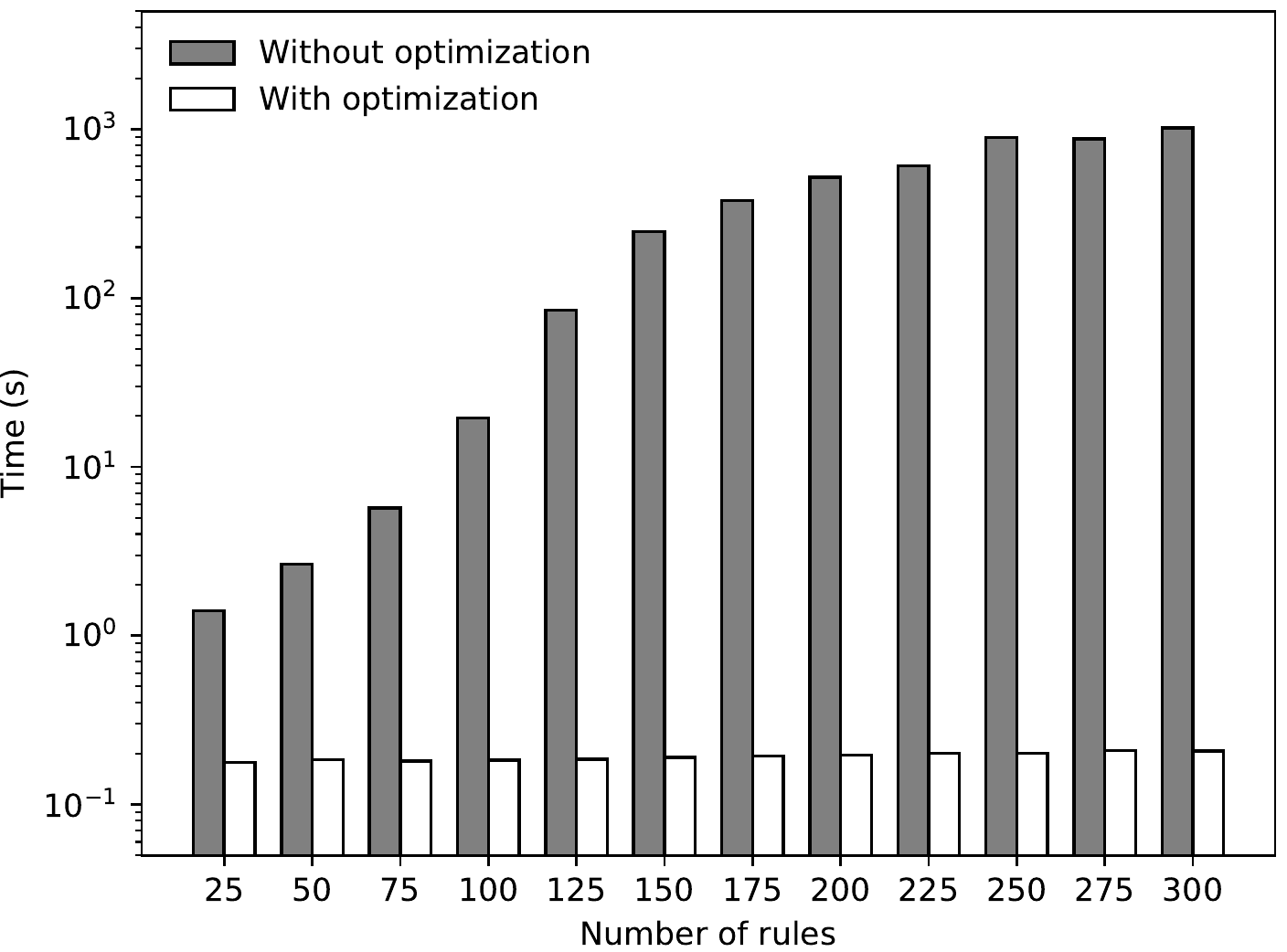}
    \caption{Performance.\\ 300 rules can be verified within less than
    one second with optimization.}
    \label{fig:performance-evaluation-of-ltl}
  \end{subfigure}
  \begin{subfigure}[t]{0.3\textwidth}
    \centering
    \includegraphics[width=\linewidth]{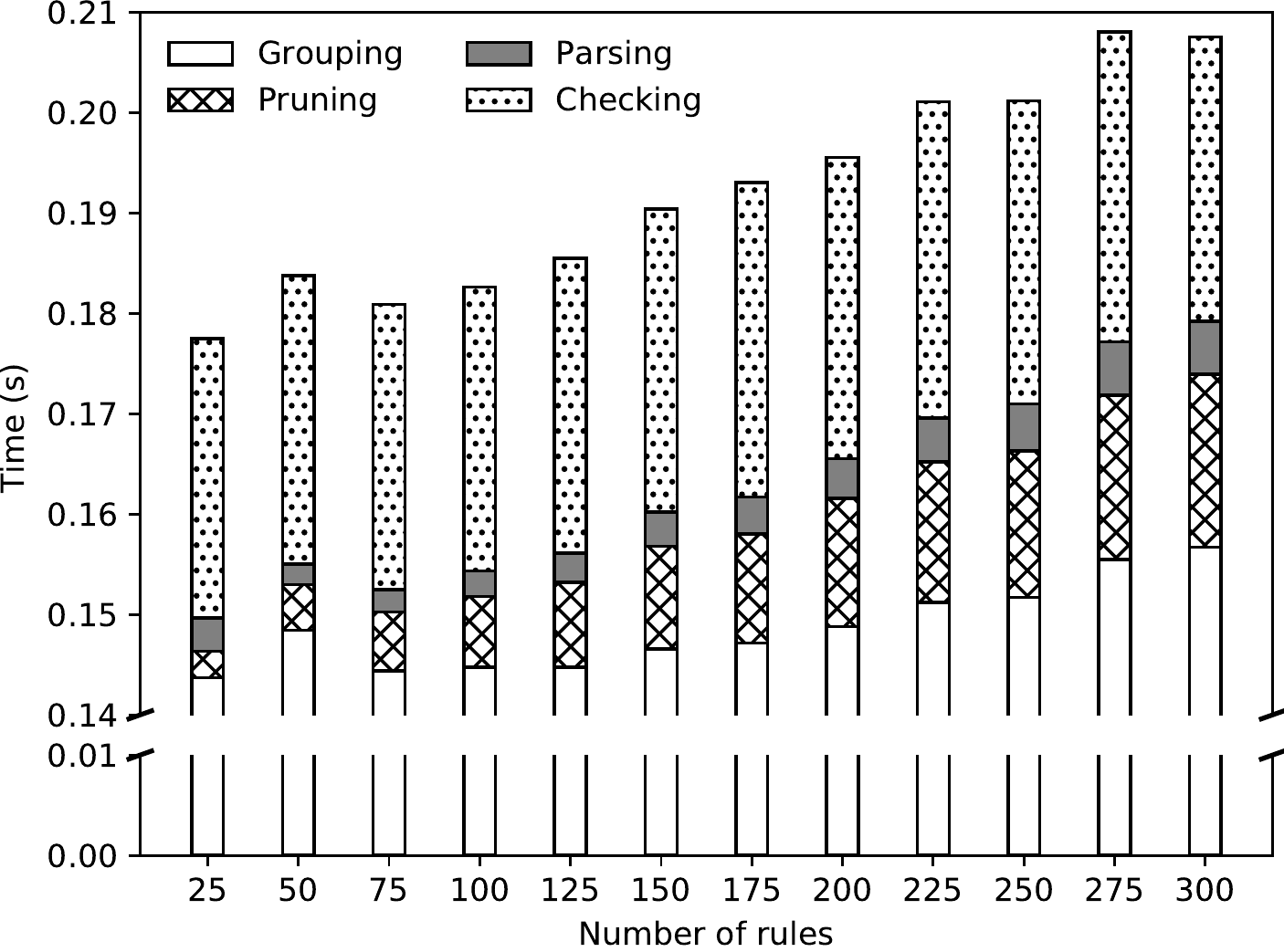}
    \caption{Optimization overhead.\\ The time needed for grouping and
    pruning only slightly increases as the number of rules increases.}
    \label{fig:optimization-overheads-of-ltl}
  \end{subfigure}
  \begin{subfigure}[t]{0.3\textwidth}
    \centering
    \includegraphics[width=\linewidth]{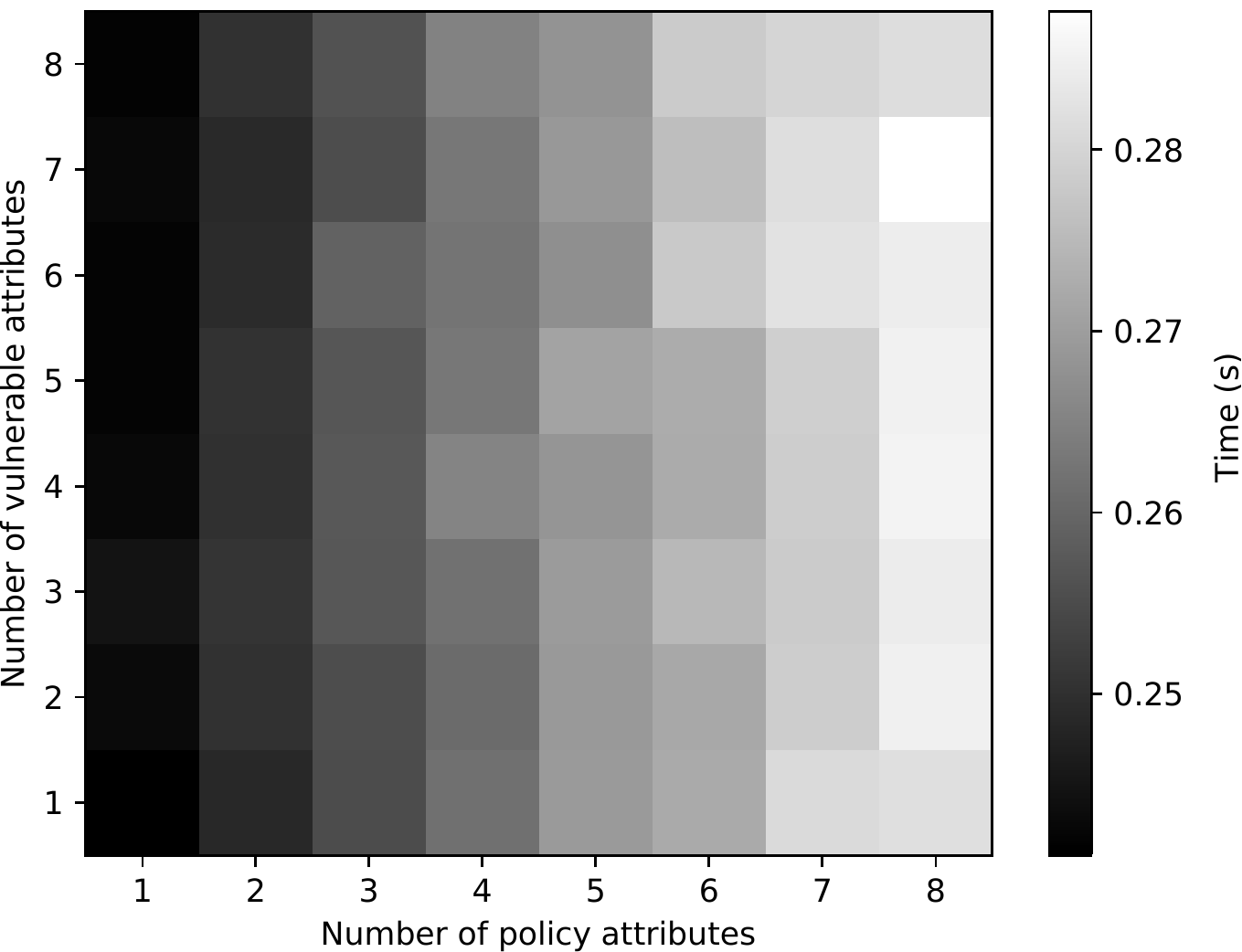}
    \caption{Effects of involved attributes.\\ Verification time
    increases slightly as the number of policy attributes increases,
    while the number of vulnerable attributes has almost no effect.}
    \label{fig:different-involved-attributes-ltl}
  \end{subfigure}
  \caption{Performance evaluation for privilege escalation in LTL}
  \label{fig:performance-evaluation-for-privilege escalation}
\end{figure*}

\paragraph{Privilege escalation.}
\name transforms security concerns 
into LTL or CTL to check if the automation rules will be
exploited.  To test the worst-case performance of \name, we randomly
build an always-TRUE specification, such that every reachable state
should be visited for verification: $G (a_i = \textbf{v} \vee
a_i \neq \textbf{v})$, where $a_i \in A$ and $\textbf{v} \in
possible(a_i)$ are both randomly chosen. We randomly select a
vulnerable attribute to simulate attacks, and each data point on the
figure represents the average of 200 experiment runs. We set a timeout
of 30 minutes (1,800 seconds) for each run.

Figure~\ref{fig:performance-evaluation-of-ltl} shows the verification
time (in a logarithmic scale) with and without our optimization
techniques. Without optimization, the verification time
increases exponentially with the number of rules. With optimization,
\name can verify 300 rules in $<1$ second. Due to space
limitations, we omit similar results of using CTL specifications.



Figure~\ref{fig:optimization-overheads-of-ltl} shows the processing
time breakdown of optimization. Pruning and grouping are the time
needed to perform respective optimizations. Checking time
represents how long NuSMV takes to return its results. Parsing
time represents the time needed to generate input files for NuSMV and
analyze its outputs. The time needed for grouping and pruning only
slightly increases as the number of rules increases.

We now evaluate the effects of different numbers of security policy
attributes and vulnerable attributes, and the result is shown in the
heatmap in Figure~\ref{fig:different-involved-attributes-ltl}.  In
  the heatmap, the execution time is represented by colors; a darker
  color represents a shorter execution time. We fix the number of
  automation rules to 500 and randomly select one attribute as
  vulnerable. 
As shown, increasing vulnerable attributes does not
greatly affect  \name's performance, because vulnerable attributes
irrelevant to policy attributes are pruned. Also, as the number of
policy attributes increases, the verification time increases slightly. The results suggest that \name
is fast enough to support complex security policies.


\begin{figure*}
  \centering
  \begin{subfigure}[t]{0.245\textwidth}
    \centering
    \includegraphics[width=\linewidth]{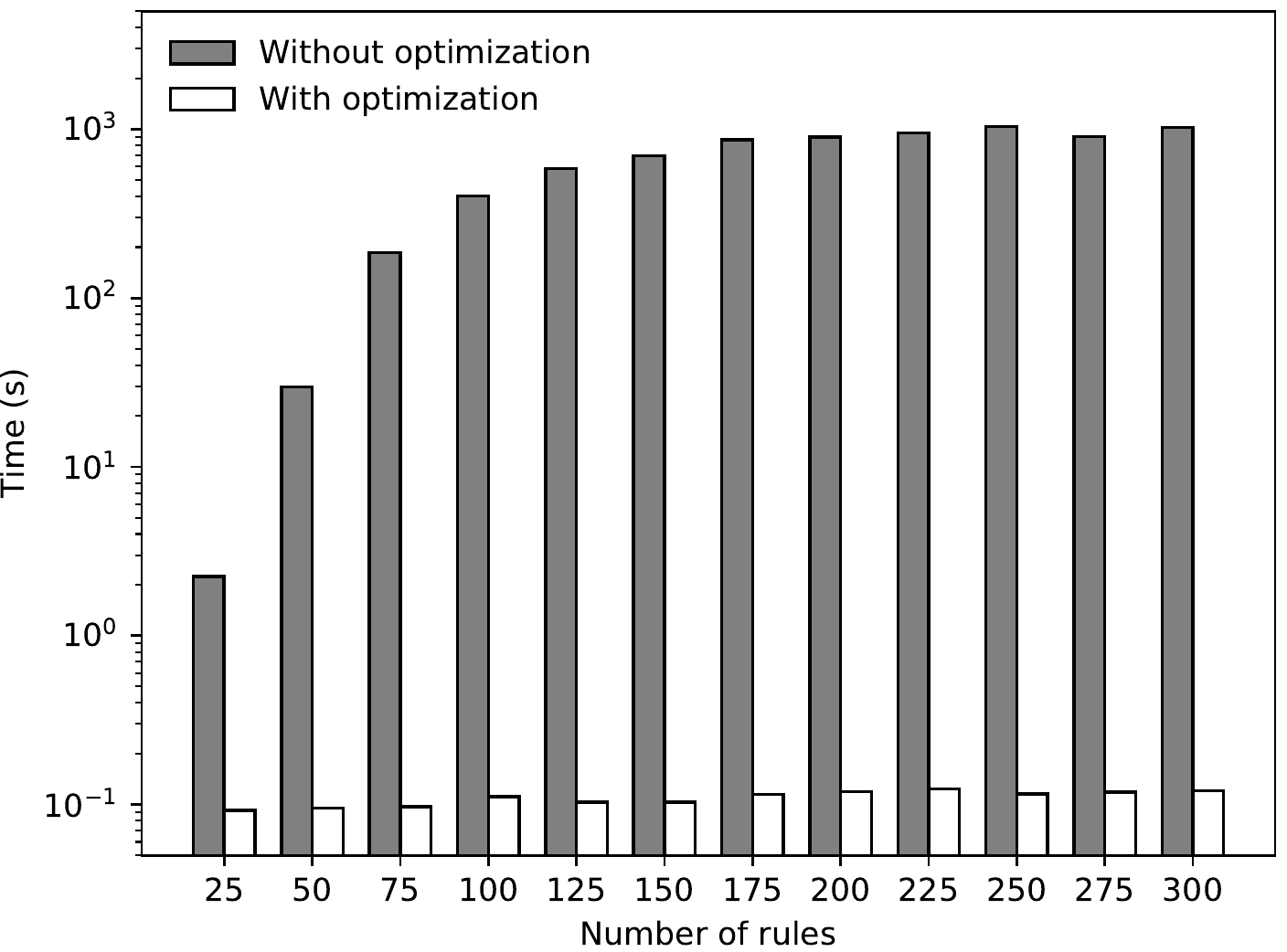}
    \caption{Performance. \\300 rules can be verified within less than
    one second with optimization.}
    \label{fig:performance-evaluation-of-privacy}
  \end{subfigure}
  \begin{subfigure}[t]{0.245\textwidth}
    \centering
    \includegraphics[width=\linewidth]{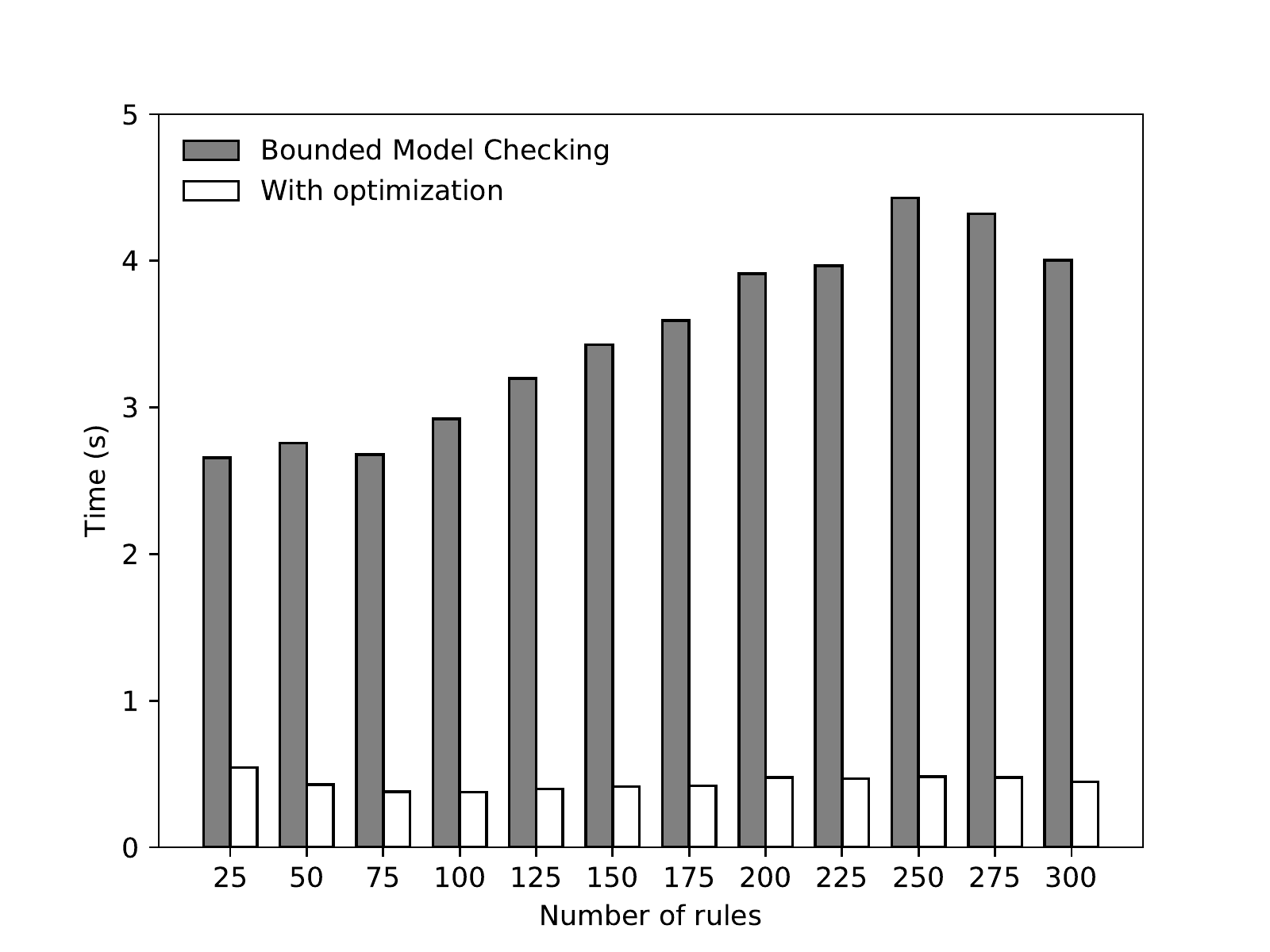}
    \caption{Performance (BMC). \\ With optimization, \name is 3-4x
      faster than BMC.}
    \label{fig:performance-evaluation-of-privacy-bmc}
  \end{subfigure}
  \begin{subfigure}[t]{0.245\textwidth}
    \centering
    \includegraphics[width=\linewidth]{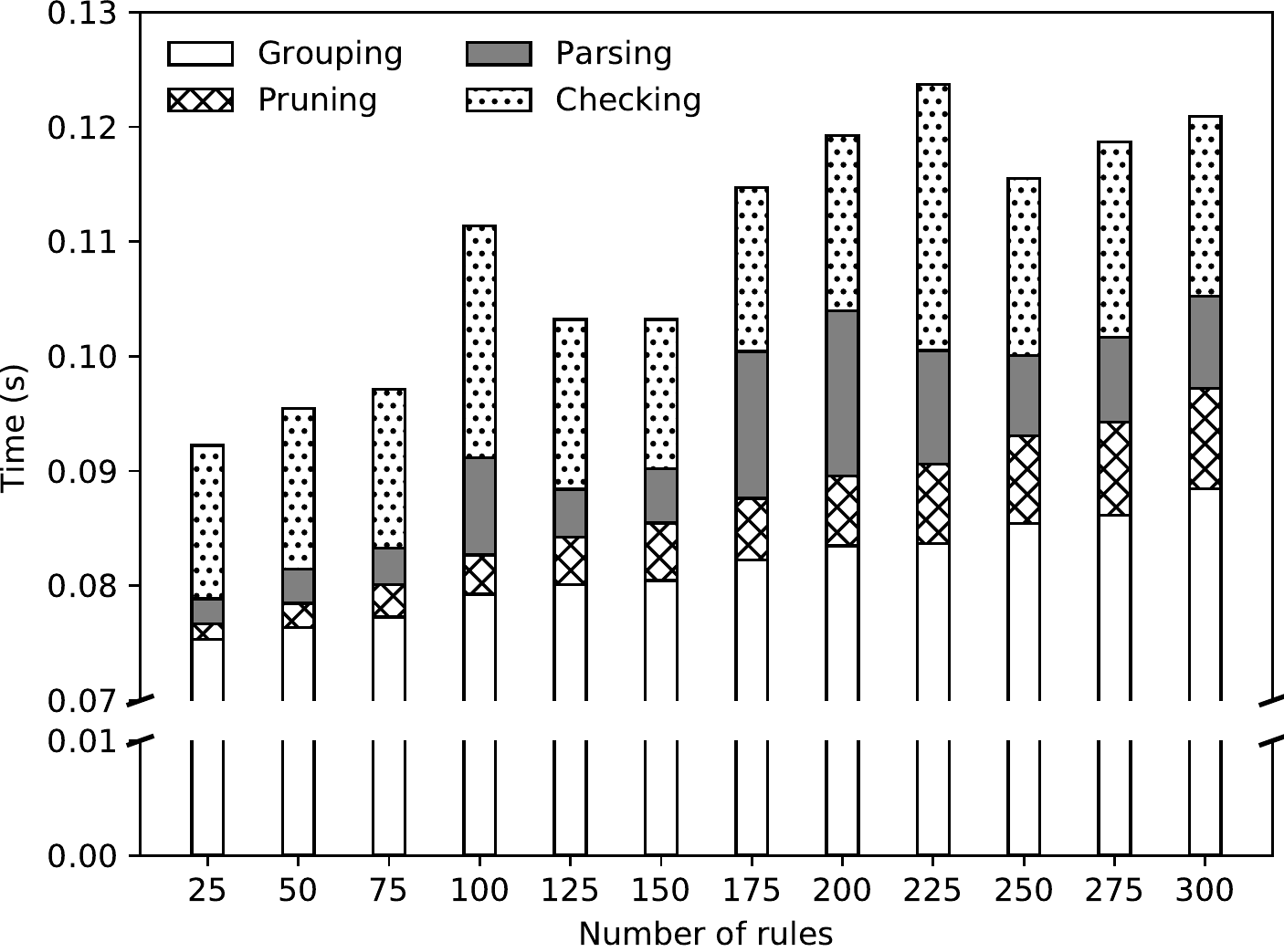}
    \caption{Optimization overhead. \\ The time grows slightly faster
    compared to \ref{fig:optimization-overheads-of-ltl} due to the use of a product
    machine.}
    \label{fig:optimization-overheads-of-privacy}
  \end{subfigure}
  \begin{subfigure}[t]{0.245\textwidth}
    \centering
    \includegraphics[width=\linewidth]{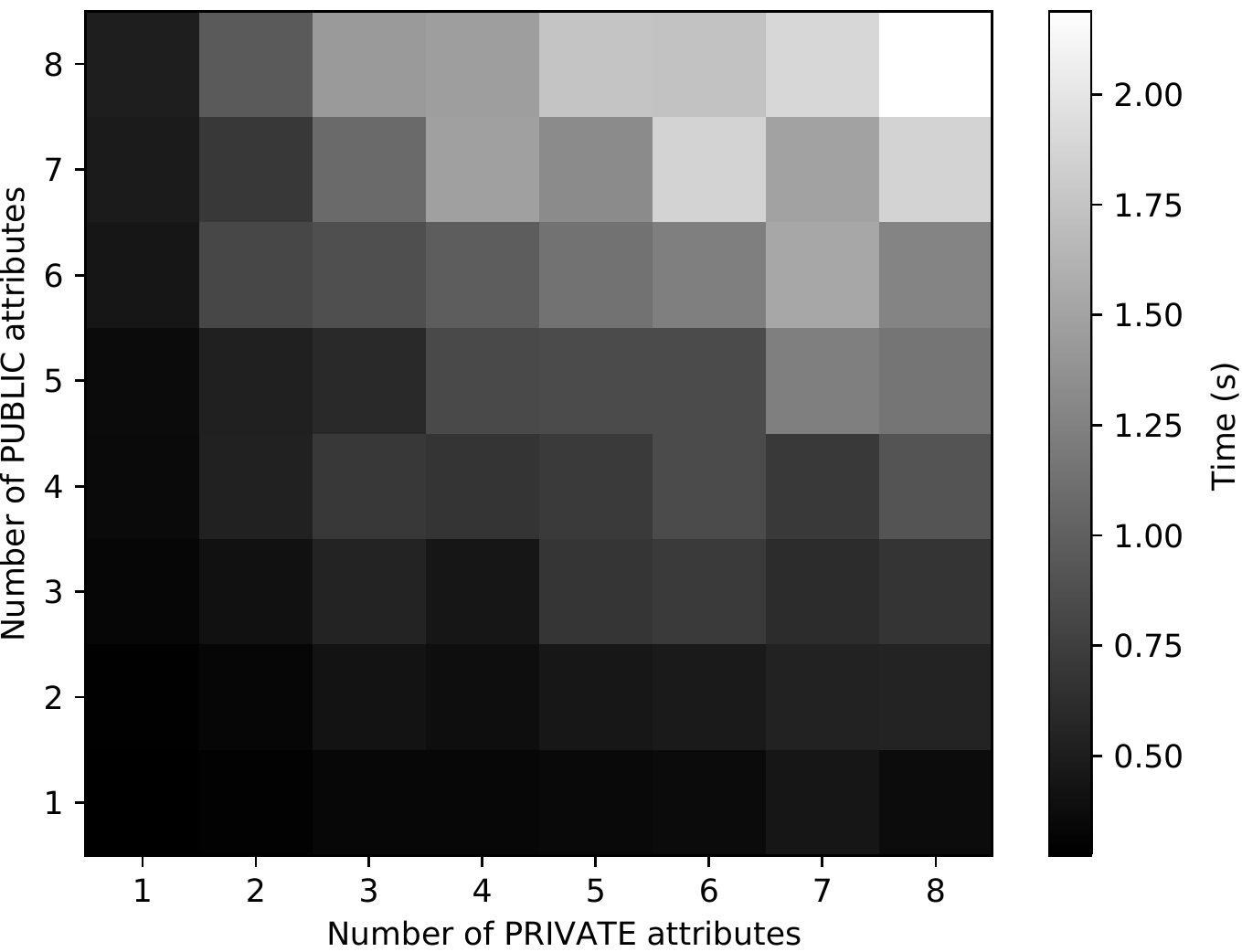}
    \caption{Different numbers of marked attributes. \\ The time
    increases only when the numbers of PUBLIC and PRIVATE variables both
    increase.}
    \label{fig:different-marked-attributes}
  \end{subfigure}
  \caption{Performance evaluation for privacy leakage}
  \label{fig:performance-evaluation-for-privacy-leakage}
\end{figure*}

\paragraph{Privacy leakage.}
Recall that users can specify what kinds of information are deemed
confidential and check whether they will be leaked to any vulnerable
attributes. To evaluate the worst-case performance, our evaluation
uses $INVAR (a_i^{FSM} = a_i^{CLONE})$ specification for verification, where $a_i^{FSM}$
and $a_i^{CLONE}$ represent the \LOW attributes $a_i \in A$ in the
original and cloned machines, respectively. The keyword $INVAR$ is a
specialized keyword provided by NuSMV to check invariance
conditions. Each time the experiment is repeated, we randomly mark an
attribute as \HIGH and another attribute as \LOW.

Figure~\ref{fig:performance-evaluation-of-privacy} shows the results
with and without our optimization. Figure~\ref{fig:performance-evaluation-of-privacy-bmc}
 shows the comparison between our optimization and Bounded Model Checking~\cite{Biere1999}
implemented by NuSMV, and Figure~\ref{fig:optimization-overheads-of-privacy}
summarizes how much time is spent during each step. As shown in these
figures, time needed for verifying privacy leakage grows much faster than privilege escalation because we use two finite state
machines to achieve pairwise comparisons. Nonetheless, these results
prove that \name is still efficient enough to detect attacks in time.


Figure~\ref{fig:different-marked-attributes} shows \name's performance
with respect to different numbers of marked attributes. The
  execution time is visualized by colors, and a darker color
  represents a shorter execution time. The experiment setup is the
same as privilege escalation. We fix the number of automation rules to
500, randomly choose different numbers of \HIGH and \LOW attributes,
and take the average over 1,000 experiments.
The figure shows that the increase of only \LOW or \HIGH attributes has
little impact on the performance of \name, because those irrelevant
attributes are pruned through optimizations. On the other hand, as the
number of both \HIGH and \LOW attributes increases, the probability of
dependency between attributes boosts, and thus fewer attributes can be
pruned. Further, since we built a product machine for pairwise
comparisons of traces, the more attributes left, the longer time needed
for verification. In this case, \name can still verify 500 automation rules
within five seconds and outperform the baseline without optimizations.

\subsection{Accuracy Evaluation}
\label{ssec:accuracy-evaluation}
We evaluate attack detection accuracy of the verification part only.
That is, we assume the devices considered vulnerable are indeed
compromised, and the environment models are accurate. Also, for simplicity,
we let $window(a_i) = possible(a_i)$.

\paragraph{Dataset.}
Our synthetic dataset shares the same channels and attributes with
those used in \S\ref{ssec:performance-evaluation}. In the dataset, the
original triggers and actions are removed, and one trigger and one
action are added for each attribute. The trigger is in the form of
$a_i = X$ and the action is $a_i \gets Y$. Both $X$ and $Y$ are
configurable parameters.

For each attack class, we conduct two experiments. These two
  experiments are designed to ensure that the ground truth is known,
  such that we can correctly identify false positives and false
  negatives.

The first  evaluates the false negative rate of our system. We first construct
an attack chain $R_1, R_2, \ldots R_l$ of length $l$ randomly, where $l$
is an integer chosen between $2$ and $8$. If the action of $R_i$ is
``$a_i \gets X$'', then the trigger of $R_{i+1}$ will be ``$a_i = X$''.
We also add another 50 rules, of which the triggers and actions  are randomly chosen.

The second experiment compares \name with previous work~\cite{Milijana2017}
that uses static analysis to show that our system can avoid false
positives in static analysis. The attack chain is similar to that of the one in the first experiment.
The only difference is that, if action of $R_i$ is ``$a_i \gets X$'', then
the trigger of $R_{i+1}$ will be ``$a_i = Y$'' and ``$X \neq Y$.'' We expect
previous work to consider such a chain to be  exploitable, since they 
statically label triggers and actions without considering user-supplied
arguments or attribute values.
Note that in this experiment we did not add additional rules to avoid accidentally
forming another attack chain.

We run each experiment 1,000 times. Below we describe the results and detail
configurations (e.g., security policies).

\paragraph{Privilege Escalation}
We mark the attributes associated with the trigger of $R_1$ as
vulnerable, and the security policy is the negation of $R_l$'s
action. Rules that are not on the chain are added with care, so that
the last action can only be triggered through the chain.
The results are as expected. \name reported the attack traces for all
the test cases in the first experiment and verified all test cases as
secure in the second experiment.

\paragraph{Privacy Leakage.}
We label the attribute associated with the trigger of $R_1$ as \HIGH,
and the attribute associated with the action of $R_l$ as \LOW.
Other attributes are all labeled OTHER.
For the first experiment, \name detects the attack chain in all 1,000
test cases, which implies that there is no false negative in our system.
For every test case in the second experiment, our system also reported no attack chain.

The experimental result demonstrates that \name has no false
negative as long as we have an accurate model. It also shows that
our approach outperforms static analysis in specific cases, as we
take runtime values into account.



\section{Discussion and Limitations}
\label{sec:discussion}
In this section, we discuss several research directions to further improve \name.

\paragraph{Adversary with partial information.}
To achieve a high level of security, \name assumes a strong attacker
who knows all of the automation rules specified by users, and thus can
also defend against a weaker attacker who knows only a subset of the
rules. An interesting future direction is to gauge an automation
system's level of security based on how much information the attacker
needs to launch a successful attack; it can be considered secure
if the attacker is required to know more than a threshold number (e.g.,
100) of automation rules.




\paragraph{Relaxing privacy notions.}
In this paper, we consider a relatively strict definition of privacy
leakage: any two states which are indistinguishable by attackers
should stay indistinguishable in the future. By satisfying this
definition, one can prevent high-valued attributes from leaking any
information to low-valued variables. This definition is relatively
strict because it does not quantify the amount of information leakage and
thus cannot differentiate a 1-bit leak from a 100-bit leak, despite the fact that
the latter is worse than the former. One interesting research
direction is to consider an analogy of an anonymity set, and quantify the
level of privacy based on the number of indistinguishable traces on
the finite state machine.  We leave it as future work to develop and
evaluate such relaxed definitions of privacy.


\paragraph{Environment modeling.}
In addition to explicit dependencies introduced by automation rules,
there are implicit dependencies enabled by proximity and environmental
changes, which can also chain automation rules together. For example, the
rule ``if the temperature is too low, turn on the heater'' seems to be
unrelated to the rule ``if the temperature is high enough, open the
window''. However, there might be a hidden relation, ``turning on the
heater will increase the temperature'', which is not explicitly specified
in the automation rules, but links the above two rules
together. Another example is that switching on bulbs can trigger light
sensors in the same room but not in other rooms. Lacking information about such implicit dependencies may cause false alarms and undetected
attacks due to missing transitions in the finite state machine.

Building an accurate environment model is challenging even with
extensive domain knowledge. \name tries to mitigate this by focusing on
the most likely scenarios in the near future and frequently
re-calibrating based on the current environment state. While
it is impossible to fully model an environment, \name can benefit
from additional information that helps reconstruct missing
transitions. For example, additional information (e.g., the location
of each device, and implicit relationships between attributes) can be
provided by the users or automatically discovered by machine learning.




\paragraph{False positives and false negatives.}
Because each counterexample reported by the model checker is indeed a
feasible attack trace with respect to the model, our scheme should
have no false positives in the ideal case. However, whether the attack
can really be conducted in the real world depends on how accurately we
model the environment. For example, if an attack can only happen when
the temperature is 100$^{\circ}\mathrm{C}$, then it is very likely to be a
false positive in the real world.

Inaccurate modeling can also cause false negatives.
One example is when a possible state does not show up in the model, and the
other is missing transitions, such as the case of implicit dependencies
discussed in the previous paragraph.

Another explanation for false negatives is when the model checker runs
out of time. This rarely happens after optimization is performed, as
our experiment results show.

\paragraph{Tuning the re-checking interval.}
Because an environment is non-trivial to model, we decided to
re-verify the automation rules as the environment changes: We first
predict how the environment will evolve during a short period of time,
and then verify the automation rules with respect to this prediction
to make sure attacks cannot be successful momentarily. After the
prediction expires, we make another round of prediction and
verification again. Because a shorter interval indicates a more
accurate environment model, a rule of thumb is setting the interval to
be the maximum time required to perform verification, which is about
1s in our implementation. This interval can be dynamically adjusted
according to the verification speed.

\paragraph{Selecting vulnerable devices.}
Using the vulnerability database is one approach to determine the
devices that are likely to be vulnerable, which is one of the inputs
to our system. There are other approaches to achieve this. For
example, users can manually select devices that require high level of
protection, and our system can evaluate whether they may be attacked
by assuming that the other devices are vulnerable.

Because the adversary's goal is to control an uncompromised device or
leak data from an uncompromised device using automation rules, it is
reasonable to consider cases where some devices are vulnerable and
some are not. It is outside the scope of this work to consider an
attacker that controls all devices, because this attacker succeeds
immediately even without trigger-action rules.

\paragraph{Limitations.}
The security of our system largely depends on external sources
such as the environment modeling, security policies and the list of
vulnerable devices. The reported 100\% accuracy in the evaluation
sections means the system will never violate the given security
policies. However, whether these policies cover all the security
aspects are outside the scope of our system.

\section{Related Work}
\label{sec:related}

The most closely related studies are discussed in
\S\ref{ssec:security-comparison}. This section reviews other related work.



\paragraph{End-user programming.}
Ur et al.~\cite{Ur2014, Ur2016} investigated the
practicality of end-user programming in the trigger-action paradigm
(``if trigger, then action'') and collected 224,590 rules shared
publicly on IFTTT. We borrowed their dataset for our large-scale
evaluation.
Mi et al.~\cite{Mi2017} conducted an empirical characterization of IFTTT,
including its ecosystem, usage and performance. They ran a
self-implemented IFTTT service in order to demystify the interaction
between a partner service and IFTTT engine. Their results show that
the current implementation of IFTTT uses a polling method for triggering,
which justifies our assumption in \S\ref{ssec:system-model}.

\paragraph{Securing service provider.}
DTAP~\cite{Fernandes2018} explored the over-privilege problem in the
IFTTT platform and found that 83\% of examined channels lack the
support of fine-grained scoping, and 75\% of the tokens are granted
more access than required to support triggers and actions. DTAP then
proposed a decentralized trigger-action platform that prevents
over-privileged service providers by using transfer tokens. DTAP
addresses an orthogonal problem, and thus can be directly combined with our
system to further enhance the overall security of automation platforms.

\paragraph{Intention mismatching.}
Huang and Cakmak~\cite{Huang2015a} investigated common inconsistencies
human users exhibit when interpreting and creating trigger-action
rules. Their results confirm the need to verify whether the actual
behaviors of user-generated rules match their high-level intentions.
By analyzing corresponding actions, TrigGen~\cite{Nandi2016}
automatically suggests missing triggers in trigger-action rules that
are composed incorrectly by users. 
Instead of fixing incorrect rules due to mismatched user
intention, our work focuses on identifying rules that might be exploited
when devices are compromised.

\paragraph{Conflict resolution.}
Ma et al.~\cite{Ma2016a} proposed a watchdog architecture for
detecting and resolving rule conflicts in the context of smart cities
based on simulation. In addition to achieving conflict detection and
resolution, CityGuard~\cite{Ma2017} further allows one to specify
safety requirements for a city.  These work focus on improving
reliability and thus do not consider the presence of adversaries.

\paragraph{SCADA and IoT.}
C{\'a}rdenas et al.~\cite{cardenas2011attacks} studied the security
issues in process control and SCADA systems and proposed to detect
stealthy attacks by incorporating the knowledge of physical
systems. They built a linear model that captures the nature of the
physical system and detect attacks using change detection
algorithms. However, building a precise model for every physical
system is impractical. Hence, our work aims to model and verify the
environment as it changes.
To compare different attack detection approaches under different
experiment settings, Urbina et al.~\cite{urbina2016limiting} proposed
a new metric to quantify attack detection algorithms and found that
stateful detection methods outperform stateless ones.

ContexIoT~\cite{Jia2017} enhances IoT applications with
context-dependent access control capabilities. ContexIoT checks
whether an action can be executed based on the current context and
asks the user to decide if the context is unclear. ContexIoT
focuses on runtime enforcement; our work is a prevention system that
eliminates potential threats before they are executed.


ProvThings~\cite{Wang2018} provides data provenance
to diagnose the trace and root cause of behavior.
However, data provenance is useful only in forensic analysis
\textit{after} an attack. While ProvThings can be extended to
support dynamic policy enforcement based on the provenance of system
events, which will notify a user when a possible attack occurs, \name
can detect an attack \textit{before} it actually happens and prevent it
by removing vulnerable rules. 

\paragraph{Temporal logics.}
Dimitrova et al.~\cite{dimitrova2012model} proposed SecLTL and SecCTL
to incorporate information flow properties into temporal logics. These
proposed logics suppose a new hide operator, which can be applied to
define privacy on a finite state machine. \name adopts a similar
definition but also incorporates temporal constraints in the context of
smart spaces. Since our definition can be viewed as a special case of
the previous one, we can express our problem using ordinary LTL and
CTL logics by building a product machine.

\section{Conclusion}
\label{sec:conclusion}
Finding the right balance between convenience and security has been a
longstanding battle, and \name is the first attempt to ease this
tension in IoT trigger-action programming.  By
transforming this programming into a model-checking problem and
formulating the security vulnerabilities into finite state machines,
\name identifies vulnerabilities among automation rules.
To overcome the
growing complexity between IoT devices and automation rules, \name
adopts pruning and grouping to ignore irrelevant devices and combine
equivalent states.  We anticipate that \name takes a step towards
securing custom automation rules in IoT for its further
advancement.





\ifCLASSOPTIONcompsoc
  \section*{Acknowledgments}
\else
  \section*{Acknowledgment}
\fi

This research was supported in part by the Ministry of Science and
Technology of Taiwan (MOST 106-2633-E-002-001 and
107-2636-E-002-005-), National Taiwan University (NTU-106R104045), and
Intel Corporation.  We gratefully thank Tiffany Hyun-Jin Kim and Matthias Schunter for their
constructive comments and suggestions in all stages of this
project. We also would like to thank Yen-Kuang Chen, Wei Jeng and Yujen Chen for
contributing to the user studies, and Shin-Min Cheng, Kai-Chung Wang,
Tony Tan and Yu-Fang Chen for their help in the app and model checking.

\ifCLASSOPTIONcaptionsoff
  \newpage
\fi



%

\begin{scriptsize}
\bibliographystyle{IEEEtranS}
\bibliography{IEEEabrv,paper}
\end{scriptsize}

\appendices
\section{Summary of IoT Channels Used}
\label{sec:channel-summary}

Table~\ref{tab:channel-summary} summarizes the IoT channels used in our experiment.

\begin{table}[h]
  \caption{42 IoT Channels Used in the Experiment}
  \label{tab:channel-summary}
  \begin{scriptsize}
    \begin{tabular}{|l|l|p{5cm}|}
      \hline
    \multicolumn{2}{|l|}{Smart phone} & Android Battery, Android Device, Android Location,
    Android Phone Call, Android Photos, Android Wear, AppZapp, Boxcar 2, Phone Call, QualityTime \\ \hline
    \multicolumn{2}{|l|}{Wearable device} & Lifelog, Nike+ \\ \hline
    \multicolumn{2}{|l|}{Smart car} & Automatic, Dash, Mojio \\ \hline
    \multirow{2}{*}{Security} & Camera & Camio, Manything \\ 
    & Door & Garageio, HomeControl Flex \\ \hline
    \multirow{10}{*}{Appliances} & Smart button & Bttn, Flic, Qblinks Qmote \\
    & Oven & GE Appliances Cooking \\
    & Water heater & GE Appliances GeoSpring \\
    & Gardening & GreenIQ, Parrot Flower Power \\
    & Printer & HP Printing \\
    & Light bulb & LIFX, Lutron Caseta Wireless, ORBneXt, Philips Hue \\
    & Sensors & Nest Protect, Netatmo Weather Station \\
    & Air conditioner & Nest Thermostat \\
    & Switch & WeMo Insight Switch \\
    & Virtual assistant & Amazon Alexa \\ \hline
    \multicolumn{2}{|l|}{Embedded system} & Adafruit \\ \hline
    \multicolumn{2}{|l|}{Web services} & Boxoh Package Tracking, ESPN, Instapush, Is It Christmas?, Printhug \\ \hline
  \end{tabular}
  \end{scriptsize}
\end{table}

\section{Dataset Collected from a Real Household}
\label{sec:YK}
We analyzed a real dataset provided by a smart home owner. The house
contains 85 connected devices, including one car, three alarms, one
camera, one energy meter, 24 light bulbs, nine motion sensors, nine
contact sensors, two smartphones, two presence sensors, 18 switches,
four thermostats, six water sensors and five weather stations. These
devices are inter-connected through about 70 automation rules. Some
rules are designed for security, such as ``if a guest arrives, turn on
camera'' and ``if nobody is at home and any motion has been detected
in the room, make siren sound.'' Some are designed for convenience and
power saving, such as ``turn off the power strip when someone is not
at home.'' The interaction between devices can be too complex to be
reasoned manually.

This dataset demonstrated that the devices used in our experiments are
similar to those being used in a real smart home. For example, one of
the most important categories are the safety and security related devices,
such as cameras, door locks, home controls and smoke sensors. We also
covered the most common sensors and actuators such as thermostats,
weather stations, light bulbs and switches.

Upon analyzing the rules in the dataset, we found some potential
attack chains that can be exploited for privilege escalation. For
instance, there is one rule ``if any family member arrives home,
switch to HOME mode,'' and another rule ``if mode transition to HOME,
disarm security cameras.'' By compromising the presence sensor, which
might be less secure, the attacker can create an illusion of someone
arrives home and finally disarm the cameras. Furthermore, when the
mode is switched to HOME, many alarms are also disabled, for example,
the rule ``if a door is open, make siren sound and send me a
message'' is executed only when current mode is not HOME. This
concludes that it is easy to form attack chains, especially when the
number of rules is large, and attack chains pose threats to real-world smart home
owners.

\section{Feasibility of Example Rules Used in This Work}
\label{sec:userstudy}
To verify the feasibility of the example rules used in this work, we
ran a user study.  We first extracted rules in
Table{~\ref{tab:chains-for-case-study}} and
Table{~\ref{tab:example-rules}}, and removed highly similar rules
($n=17$, hereafter: Group {\name}). To mix them with an equal
  number of rules sampled from the IFTTT website (Group Reference), we
  then used the devices shown in Figure{~\ref{fig:home-overview}}
(e.g., smart lock) as keywords to collect the first 15 rules
  displayed on the returned page for each device.  After eliminating
  duplicates and highly similar rules (e.g., same rules for different
  brands) from the obtained 90 rules, we randomly sampled 17 rules and
  created a survey with 34 rules in total ($n=17$ for Group {\name}
  and Group Reference each). There were no overlapping rules in these
  two groups.


  We created a question for each rule, and the question asked a
    participant to rate his/her willingness to use the rule in a
    Likert scale from 1 (totally disagree) to 5 (totally agree). The
    rules were displayed in a randomized order to avoid the sequential
    effect.


  We recruited 108 participants who have prior experience with
    IFTTT rules using Amazon Mechanical Turk.  The survey took
    approximately 10 minutes and we paid \$2.00 for all participants.
    After eliminating those who did not have sufficient background
    knowledge on IFTTT rules, we eventually obtained 79 valid
    responses. Most participants (66, 83.5\%) reported to own 1-5
    smart home devices, 9 (11.4\%) reported to own 6-10 smart home
    devices, and 4 (4.7\%) participants reported to own 11-20
    devices. Among the 79 valid responses, the top three popular
    devices are smart TV (n=36, 45.6\%), smart thermostat (n=16,
    20.3\%), and smart speaker (n=11, 13.92\%). The most popular hubs
    are Amazon Alexa (47 out of 79) Google assistant (25 out of 79),
    and Samsung SmartThings (22 out of 79). 19\% of the participants
    reported themselves to be extremely experienced and 53.2\% report
    somewhat experienced with IFTTT. In terms of participant
    demographics, 36.7\% of the participants are females, and the
    majority of the participants are between ages 25 and 34 (54.4\%,
    n=43) and between ages 35 and 44 (24.1\%, n=19). 38\%, 22.8\% and
    15.2\% of the 79 participants reported to hold a highest degree in
    bachelor degree, master degree, and some college, respectively. 

  Based on their responses, we examined the willingness level of
  individual users as well as combinations of users. We used a
    T-test to check if these two groups of rules have significant mean
    difference, and we found some significant difference ($t(78) =
    -6.04$, $p < .0001$). This result suggests that users are more
    willing to use the rules in Group Reference than Group {\name},
    which is reasonable because Group {\name} intentionally includes
    potential attack chains for demonstration purposes. Based on this
    result, we are unable to show that people may adopt the rules in
    Group {\name} in practice. Hence, we performed further
    analysis and found that 78.5\%, 45.6\%, 67.7\% and 70.9\% of
    participants are willing (with a Likert score of 3 or higher) to
    use the four chained rules in
    Table{~\ref{tab:chains-for-case-study}}, respectively. This result
    shows that a large percentage of users may adopt rules that lead to unintentional
    chained effects in practice.

  In addition, IFTTT rules may be created by multiple users that share the
  same living/working space. To check the willingness to use with
    multiple users (say a 3-person family), we enumerated all possible
    combinations of three participants and examined their aggregated
    willingness by taking the maximum (out of the three) of the Likert
    scale values. We found that 78\% of the combinations of the 3
    participants would be willing (with a Likert score of 3 or higher)
    to use all the rules in Group {\name}, while 91\% of the
    combinations of the three participants will be willing to use all
    the rules in Group Reference. In short, this user study results
    imply that while the rules in Group {\name} are less appealing
    than Group Reference to individual users, the likelihood of three
    or more users in a group to use all the rules in Group {\name}
    remains high, implying the practicality of these rules for the
    real-world scenarios.


%








\end{document}